\documentclass[usenatbib]{mnras}

\usepackage{newtxtext,newtxmath}
\usepackage[T1]{fontenc}
\usepackage{ae,aecompl}

\usepackage{mathtools}
\usepackage{siunitx}
\usepackage{multirow,tabularx}
\usepackage{amsmath}
\usepackage{graphicx}
\usepackage{comment}
\usepackage[normalem]{ulem}
\usepackage{BibDef}
\usepackage[dvipsnames]{xcolor}
\usepackage{hyperref}
\hypersetup{
	colorlinks=true,        % false: boxed links; true: coloured links
	linkcolor=blue,         % colour of internal links
	citecolor=blue,         % colour of links to bibliography
}
\newcommand{\msun}{~\rm M_{\large \odot}}
\newcommand{\ud}{~{\rm d}}

\newcommand{\galpy}{{\textsc{galpy}}}

\title[MP evolution in realistic galaxy models]{Dynamical evolution of massive perturbers in realistic multi-component galaxy models I: implementation and validation}

\author[M. Bonetti et al.]{Matteo Bonetti,$^{1,2,3}$ Elisa Bortolas,$^{1,4}$ Alessandro Lupi$^{1,5}$ and Massimo Dotti$^{1,2,3}$
\\
$^1$Dipartimento di Fisica ``G. Occhialini'', Universit\`a degli Studi di Milano-Bicocca, Piazza della Scienza 3, I-20126 Milano, Italy\\
$^2$INFN, Sezione di Milano-Bicocca, Piazza della Scienza 3, I-20126 Milano, Italy\\
$^3$INAF, Osservatorio Astronomico di Brera, Via E. Bianchi 46, I-23807, Merate, Italy\\
$^4$Center for Theoretical Astrophysics and Cosmology, Institute for Computational Science, University of Z\"urich, Winterthurerstrasse\\ \ 190, CH-8057 Z\"urich, Switzerland\\
$^5$Scuola Normale Superiore, Piazza dei Cavalieri 7, I-56126 Pisa, Italy\\
}

\date{Accepted 2021 January 22. Received 2021 January 19; in original form 2020 October 16}

\pubyear{2021}

\begin{document}
\label{firstpage}
\pagerange{\pageref{firstpage}--\pageref{lastpage}}
\maketitle

%%%%%%%%%%%%%%%%%%%%%
\begin{abstract}
  Galaxies are self-gravitating structures composed by several components encompassing spherical, axial and triaxial symmetry. 
  Although real systems feature heterogeneous components whose properties are intimately connected, semi-analytical approaches often exploit the linearity of the Poisson's equation to represent the potential and mass distribution of a multi-component galaxy as the sum of the individual components.
  In this work, we expand the semi-analytical framework developed in Bonetti et al. (2020)
  by including both a detailed implementation of the gravitational potential of exponential disc (modelled with a ${\rm sech}^2$ and an exponential vertical profile) and an accurate prescription for the dynamical friction experienced by massive perturbers in composite galaxy models featuring rotating disc structures. Such improvements allow us to evolve arbitrary orbits either within or outside the galactic disc plane.
  We validate the results obtained by our numerical model against public semi-analytical codes as well as full N-body simulations, finding that our model is in excellent agreement to the codes it is compared with.
  The ability to reproduce the relevant physical processes responsible for the evolution of massive perturber orbits and its computational efficiency make our framework perfectly suited for large parameter-space exploration studies.
\end{abstract}
%%%%%%%%%%%%%%%%%%%%%


%%%%%%%%%%%%%%%%%%%%%%%%%%%%%%%%%%%%%%%%%%%%%%%%%%%%%%%%%%%%%%%%%%%%%%%%%%%%%%%%%%%%%
%%%%%%%%%%%%%%%%% BODY OF PAPER %%%%%%%%%%%%%%%%%%
\section{Introduction}
\label{sec:intro}
 
In the last decades, more and more interest has grown towards the modelling of orbital motions within the potential of a galaxy \citep[e.g.][]{galpy,Boubert2020,Granados2021}.
This has to be attributed to the advent of missions like Gaia \citep{gaia1, gaia2}, which  is tracking the orbits of stars and stellar clusters sailing though the  Galactic potential, and will release an unprecedented catalogue of Galactic stellar orbits. In addition, gravitational wave observatories such as LISA \citep{LISA2017} and PTA \citep{Verbiest2016} will probe massive black hole (MBH) mergers along the cosmic time; in order to interpret the rates of such mergers, one needs to model the inspiral phase of an intruder MBH within the potential of the host galaxy, down to the scale at which it will form a binary with another black hole that may already sit in the centre \citep{Begelman1980}.

A detailed modelling of galactic orbits can only be achieved with a sufficiently good description of the associated galactic potential wells; in addition, if one wants to accurately follow the orbit of a massive perturber (MP, i.e. an object whose mass is significantly larger than the characteristic stellar mass), it is crucial to further account for the effect of dynamical friction \citep[DF,][]{Chandra43,BT}.

At the present time, state-of-the art numerical simulations with sub-grid recipes for unresolved physical processes promise to be the best tool to properly model the orbital evolution of stars and MPs within galaxies \citep[e.g.][]{Tremmel2015, Pfister2017, Pfister2019}. However, they cannot be adopted when one needs to perform a throughout exploration of the parameter space as they are typically very computationally expensive. For this, different tools, such as semi-analytical approaches, may be preferred owing to their much lower computational cost, that comes at the expense of some precision.

In semi-analytical models, the motion of a test particle subject to the galactic potential can be followed more and more accurately as the description of the potential associated to each galactic component (the dark matter halo, disc, bulge etc.) gets closer to the one of real galaxies. While the spherically symmetric galactic components can be approximated reasonably well by well-known potential-density pairs featuring a simple analytical form \citep[e.g.][]{Navarro1997, Plummer1911, Dehnen1993}, the same cannot be said for less idealized galactic structures such as discs. In fact, their light distribution is often well described by an exponential disc density profile, which unfortunately does not admit a close analytical form for the related potential, meaning that its computation has to be performed numerically and the associated computational cost may be relatively large. For this, less physically motivated disc potentials featuring a simpler, close analytical form are often adopted in the literature, even if their approximation of the potential of a physical galactic disc may be relatively poor.\\

Even neglecting the aforementioned issues, a proper description of the galactic potential in semi-analytical models is not enough when the target particle is a MP, as DF may significantly affect the evolution of its orbit. DF arises as a response of the background to the perturbing mass, and it typically results in a deceleration of the latter, which gradually sinks towards the centre of its host system. Its standard semi-analytical description grounds its basis on the seminal work of \citet{Chandra43}, who first analytically modelled the motion of an MP in an infinite and homogeneous stellar background.\footnote{Under similar assumptions, \citet{Ostriker1999} describel the analogous phenomenon within a gaseous  medium.} 
In spite of its simplicity,  \citeauthor{Chandra43}'s approach has been proven to work remarkably well in many scenarios, sometimes far beyond the expected regime of applicability \citep[e.g.][]{White1983},
and despite missing the physics of resonant interactions between the MP and the host system \citep{Tremaine1984, Weinberg1986}. Owing to its success, the semi-analytical treatment for DF has seen a substantial development throughout the last decades, as attested by the vast piece of  recent literature  focusing on improving  Chandrasekhar's prescriptions and their agreement with N-body integrations \citep[e.g.][]{Hashimoto2003, Just2005, Just2011, Arca-Sedda2014, Read2006, Petts2015, Petts2016}.  
However, the  work in this direction has been limited to very idealized, spherically symmetric and isotropic systems, which can only describe a narrow range of galaxy environments; in fact, astrophysical systems feature much more complexity than this: an extension of the semi-analytical DF treatment to composite  systems, possibly featuring disky components with rotational support, would thus greatly enhance the realm of applicability of semi-analytical DF. 

N-body experiments already  addressed the effect of DF in a composite, rotationally supported environment; they highlighted that DF acting in rotationally supported systems induces the circularization of a MP initially on a  prograde, eccentric orbit, and reverses the angular momentum of counter-rotating MPs, then again promoting circularization \citep{Dotti06,  Callegari11, Fiacconi13};  this effect appears to be independent of the nature of the background \citep{Dotti07}. However, such past studies never attempted a quantitative modelling of the phenomenon. Only very recently, \citet{Bonetti2020} successfully modelled for the first time the DF-induced ``drag towards circular co-rotation'' in a semi-analytical fashion in a multi-component galaxy that featured a dark matter halo, a stellar disk and a bulge; their prescription showed remarkable agreement with N-body simulations down to the circularization phase.
However, their study was limited to on-plane orbits, and their approach can be adopted only before circularization occurs.

Here, we expand the work presented in \citet{Bonetti2020} by improving the DF treatment in rotating discs with the development of a more physical DF prescription. Such prescription features no arbitrary tunable parameters and through the employment of a more realistic (but still isotropic) disc velocity distribution function we extend its validity also to circular orbits. Further to DF treatment, we also provide a completely revisited implementation for the acceleration evaluation due to an exponential disc mass distribution, now allowing arbitrary orbits and not only equatorial motion.

On a longer time-span, this work is intended to be the first of a series aiming at providing a semi-analytical computational framework that encodes the most realistic as possible dynamics of MPs in multi-component galaxies, but at the same time still featuring a high computational efficiency, able to guarantee vast and systematic parameter space explorations in terms of diverse galactic profiles and MP initial properties/trajectories.

The paper is organised as follows: in Section~\ref{sec:method} we describe the improvements brought to the framework developed in \citet{Bonetti2020}; in Section~\ref{sec:results} we validate our updated setup against existing semi-analytical codes and full N-body simulations featuring the same initial conditions. Finally, in Section~\ref{sec:discussion} we draw our conclusions.

%%%%%%%%%%%%%%%%%%%%%%%%%%%%%%%%%%%%%%%%%%%%%%%%%%%%%%%%%%%%%%%%%%%%%%%%%%%%%%%%%%%%%
\section{Method}
\label{sec:method}

As in \citet{Bonetti2020}, we consider a multi-component galaxy model comprising a Navarro-Frenk-White \citep[NFW,][]{Navarro1997} profile to model the dark matter (DM) halo, an Hernquist \citep{Hernquist1990} profile to describe a compact stellar bulge and a thick exponential disc profile to characterise the galactic disc. On top of that we model the effect of DF on the motion of massive perturbers through the employment of dissipative forces. 

In this section we detail the procedure we followed to i) evaluate the potential and accelerations generated by a thick exponential disc in the general case, ii) discuss possible improvement for the DF force in spherically symmetric systems and iii) extend the validity of our prescription for DF in rotating discs  to nearly circular orbits.   

%%%%%%%%%%%%%%%%%%%%%%%%%%%%%%%%%%%%%%%%%
\subsection{Exponential disc conservative dynamics}
\label{sec:exp_disc}

We consider a finite-thickness exponential disc whose mass density is described by

%%%%%%%%%%%%%%%
\begin{equation}\label{eq:dens_disk}
    \rho_d(R,z) = \dfrac{M_d}{4\pi R_d^2 z_d} {\rm e}^{-R/R_d} {\rm sech}^2\left(\dfrac{z}{z_d}\right),\\
\end{equation}
%%%%%%%%%%%%%%%
where $M_d, R_d$ are the total mass and scale radius of the disc, while $z_d$ is a parameter shaping the profile along the $z$ direction.
Such profile is our choice for the modelling of general galactic discs, since it can well reproduce the light profiles of real galaxies and it is physically motivated, being the vertical distribution obtained for an isothermal self-gravitating disc \citep{Spitzer42}. For these reasons, it is usually employed to generate the initial conditions of N-body simulations \citep[see e.g.][]{yurin14}, allowing for a validation of the results of our semi-analytical prescriptions against N-body runs. 
Nevertheless, the double exponential disc, i.e. a disc with vertical density profile decaying as ${\rm e}^{-|z|/z_d}$, is widely used and well reproduces the vertical profiles of edge-on disc galaxies \citep[e.g.][]{deGrijs97}. In order to have a proper comparison, we also implemented this second profile in our framework.\footnote{The double exponential disc also does not admit any closed form for the gravitational potential, but requires numerical integration as given in equation~\eqref{eq:disk_pot}, where the sole difference is in the function $I(k)$ given by equation~A9 of \citet{Kuijken1989}.}

In general, to obtain the gravitational potential from a given density distribution, one has to solve the Poisson's equation. Practically, when deviating from spherical symmetry, a complete analytical expression can be found only in a handful of cases  and unfortunately the thick exponential case does not belong to those. Therefore, to obtain the gravitational potential (and consequently the accelerations) of such mass distribution at arbitrary $(R,z)$ pairs, we necessarily need to recur to numerical methods. Unfortunately, the general solution of the Poisson's equation that can be obtained via Green's function results in multidimensional integrals, implying that obtaining  the solution  can be pretty expensive in computational terms. Thus, if possible, it is advisable to reduce the expressions as much as possible before recurring to numerical integration.

It turns out that, when considering an axi-symmetric density distribution, the Poisson's equation can be solved in terms of a Hankel's transform \citep[see e.g.][]{Kuijken1989}. This approach allows us to write the gravitational potential as a 1-dimensional integral (see Appendix~\ref{sec:appA} for a complete derivation), given by 

%%%%%%%%%%%%%%%%
\begin{equation}\label{eq:disk_pot}
    \phi(R,z) = - \dfrac{G M_d}{2 R_d^3 z_d} \int_0^\infty \ud k J_0(k R) \dfrac{I_z(k)}{\left(R_d^{-2}+k^2\right)^{3/2}},
\end{equation}
%\frac{1}{R_d^2} 
%%%%%%%%%%%%%%%%
where $J_0$ is the Bessel function of the first kind,  and $I_z(k)$ is a function depending on both $k$ and $z$ through the Gauss hypergeometric function $_2F_1$ (see Appendix~\ref{sec:appA} and \ref{sec:appB}).

The radial and vertical accelerations can be readily obtained by taking the (negative) derivative of equation~(\ref{eq:disk_pot}) with respect to $R$ and $z$ and then performing the integration over the $k$ variable 

%%%%%%%%%%%%%%%%
\begin{align}\label{eq:acc_R}
    a_R = -\dfrac{\partial \phi}{\partial R} &= - \dfrac{G M_d}{2 R_d^3 z_d} \int_0^\infty \ud k k J_1(k R) \dfrac{I_z(k)}{\left(R_d^{-2}+k^2\right)^{3/2}}, \\
    a_z = -\dfrac{\partial \phi}{\partial z} &= - \dfrac{G M_d}{2 R_d^3 z_d} \int_0^\infty \ud k J_0(k R) \dfrac{-\partial_z I_z(k)}{\left(R_d^{-2}+k^2\right)^{3/2}}.
    \label{eq:acc_z}
\end{align}
%%%%%%%%%%%%%%%%
Note that the integration of Bessel functions could be problematic given their  highly oscillatory behaviour. Nevertheless, if appropriate quadrature techniques are employed,  the integrals can be computed quite efficiently using a few hundreds of points, still guaranteeing a good level of accuracy (see Appendix~\ref{sec:appA}).

%%%%%%%%%%%%%%%%%%%%%%%%%%%%%%%%%%%%%%%%%
\subsection{DF in spherical profiles}

DF, though not dissipative in nature, in the context of semi-analytical calculations is usually modelled as a drag force acting opposite to the velocity of a MP. Despite the effort to improve the original derivation by \citet{Chandra43} \citep[see e.g.][and reference therein]{Tremaine1984,Weinberg1986,Mulder1983,Colpi1999} the rather simple original expression, i.e.
 
%%%%%%%%%%%%%%%%%%
\begin{equation}\label{eq:DF_sph}
    \mathbf{a}_{\rm df} = -2\pi G^2 \ln(1+\Lambda^2) m_p \rho \left({\rm erf}(X) - \dfrac{2 X{\rm e}^{-X^2}}{\sqrt{\pi}}\right) \dfrac{\mathbf{v}_{p}}{|\mathbf{v}_{p}|^3},
\end{equation}
%%%%%%%%%%%%%%%%%%
is still widely used, mostly because of its simplicity.
In the above formula, $m_p$ and $\mathbf{v}_p$ are the MP mass and velocity, $\rho$ is the background density, $\Lambda = p_{\rm max}/p_{\rm min}$ is the ratio between the maximum and minimum impact parameter, while $X=v_p/(\sqrt{2}\sigma)$ is the ratio of the MP velocity over the velocity dispersion. Though systematically employed, equation~\eqref{eq:DF_sph} is subjected to a number of assumptions that are not always justified in realistic galactic contexts, i.e.:
%%%%%%%%%%%%%%%%%%
\begin{itemize}
    \item[1.] the motion of scattering particles with large impact parameters is rectilinear;
    \item[2.] the background density is homogeneous and isotropic;
    \item[3.] the velocity distribution function is isotropic and Maxwellian with constant dispersion $\sigma$;
    \item[4.] the maximum impact parameter is chosen equal to a maximum scale-length of the system, while $p_{\rm min}$ is kept fixed for all possible encounters and equal to a characteristic scale length (or, in the case of point-like perturbers, the radius below which a $90^\circ$ deflection occurs). Specifically, this last assumption implies that only encounters with velocity lower than the MP velocity contribute to the deceleration.
\end{itemize}
%%%%%%%%%%%%%%%%%%

Despite the above assumptions are clearly not fully compatible with realistic galaxies, the introduction of some a-posteriori modifications (mostly affecting points 2 and 3 above) to equation~\eqref{eq:DF_sph} allows to reproduce the results from N-body simulations with reasonable accuracy. In particular, when limiting the DF treatment to spherically symmetric profiles (which model to a fairly good approximation elliptical galaxies or galaxy bulges), the adoption of the position-dependent mass density, velocity dispersion and minimum and maximum impact parameters substantially improves the agreement between semi-analytical models and full $N$-body simulations \citep[see e.g.][]{Hashimoto2003, Just2005, Just2011, Petts2015, Petts2016}. 
Following \citet{Bonetti2020}, we implement equation~\eqref{eq:DF_sph} adopting the position-dependent velocity dispersion $\sigma(r)$ and mass density $\rho(r)$, and the following expressions for the maximum and minimum impact parameters: 
%%%%%%%%%%%%%%%%%%
\begin{align}\label{eq:finesse}
    p_{\rm max} &= r/\gamma, \nonumber \\
    p_{\rm min} &= \max\left(\dfrac{G m_p}{v_p^2+\sigma(r)^2}, D_p\right), \nonumber \\ 
    \gamma &= - \dfrac{\ud \ln \rho}{\ud \ln r},
\end{align}
%%%%%%%%%%%%%%%%%%
with $\gamma$ denoting the logarithmic slope of the density profile, $r$ the radial coordinate and $D_p$ the physical radius of the MP (which can be set to zero for the case of an MBH).  

Even with the above modifications, the acceleration computed with the functional form of equation~\eqref{eq:DF_sph} considers only the contribution of scattering particles moving slower than the MP. Although this does not represent a serious issue for the majority of stellar systems, the deceleration described by equation~\eqref{eq:DF_sph} could severely underestimate the real DF when the number of background objects moving faster than the MP is large. 
Specifically, \citet{Antonini2012} investigated the DF experienced by a MP in galactic cores dominated by a central MBH and they found that, 
when the slope of the density profile is $\gamma \lesssim 1$, the contribution of the so-called ``fast-moving stars'' can become comparable to or even larger than that of slow moving ones. 

In order to take into account such fast population, one needs to drop some assumptions and consider the general Chandrasekhar formula \citep[see equations 25 and 26 of][]{Chandra43}, that despite being more general, usually does not allow for an analytical form, i.e.\footnote{The integral over the relative velocity $J(v_p,v_\star,p_{\rm max})$ actually has an analytical form. It is usually quoted in an approximate form that is well behaved everywhere except when $v_p$ and $v_\star$ are close \citep[see e.g. equation 7 of][]{Antonini2012}. For completeness we report the full expression in Appendix \ref{sec:appC}.}
%%%%%%%%%%%%%%%%%%
\begin{align}\label{eq:DF_general}
    \mathbf{a}_{\rm df} = -4\pi G^2 m_p & \rho(r) \dfrac{\mathbf{v}_p}{|\mathbf{v}_p|^3} \times \nonumber \\
    & \times \int_0^{v_{\rm esc}} \dfrac{J(v_p,v_\star,p_{\rm max})}{8\pi v_\star} 4\pi v_\star^2 f(v_\star) \ud v_\star,
\end{align}
%%%%%%%%%%%%%%%%%%
%%%%%%%%%%%%%%%%%%
\begin{align} \label{eq:DF_general_J}
    &J(v_p,v_\star,p_{\rm max}) = \int_{|v_p-v_\star|}^{v_p+v_\star} \ud V \left(1+\dfrac{v_p^2-v_\star^2}{V^2}\right) \ln\left(1+\dfrac{p_{\rm max}^2 V^4}{G^2 m_p^2}\right).
\end{align}
%%%%%%%%%%%%%%%%%%

Equation~\eqref{eq:DF_general} inevitably implies a higher computational burden due to the unavoidable numerical computation of the integral over the distribution function. Given that in the vast majority of astrophysical situations the inclusion of fast moving stars represents only a minor correction (but see Section~\ref{sec:discussion} for some caveats of our choice), in this work we adopt the simplified expression considering only slow moving stars, together with the modifications introduced by different authors that improve its range of applicability.

%%%%%%%%%%%%%%%%%%%%%%%%%%%%%%%%%%%%%%%%%
\subsection{DF in rotating discs}

\citet{Bonetti2020} derived a prescription to describe the DF acceleration acting on a MP determined by a rotating disc,

%%%%%%%%%%%%%%%%%%
\begin{equation}\label{eq:DF_disk22}
    \mathbf{a}_{\rm df,disc} = -A_{\rm disc} 2\pi G^2 \ln(1+\Lambda^2) m_p \rho_d(R,0) \dfrac{\mathbf{v}_p-\mathbf{v}_c(R)}{|\mathbf{v}_p-\mathbf{v}_c(R)|^3},
\end{equation}
%%%%%%%%%%%%%%%%%%
where $\mathbf{v}_p,m_p$ are the velocity and mass of the MP, $\mathbf{v}_c(R)$ is the circular velocity at radius $R$ determined by the total gravitational potential, while $\Lambda = p_{\rm max, d}/p_{\rm min, d}$, that enters an effective Coulomb logarithm, is chosen as the ratio of the maximum impact parameter, taken equal to the $z_d$ parameter of the disc, and a minimum one given by 
%%%%%%%%%%%%%%%%
\begin{align}\label{eq:pmin_circ}
   p_{\rm min, d} &= \dfrac{G m_p}{v_{\rm rel}^2 + 0.01 v_c^2},\nonumber\\
   v_{\rm rel} &= |\mathbf{v}_p-\mathbf{v}_c|.
\end{align}
%%%%%%%%%%%%%%%%
Finally, the factor $A_{\rm disc}$ is a tunable constant set by comparing the semi-analytical orbital evolution of the MP with that obtained in an N-body simulation. \citet{Bonetti2020} found that this factor is of the order of unity, enforcing that their modeling catches the main physics of DF in discs.

The main limitation of this prescription lies on the dependence of the DF acceleration on the relative velocity with respect to circular velocity. When the MP velocity approaches the circular one, equation~\eqref{eq:DF_disk22} diverges, thus the orbital decay cannot be followed further. 

To overcome this limitation, here we consider a less idealized distribution function for the stars in the disc. Specifically, we replace the delta function with an isotropic Gaussian centered around the local rotational velocity.
This choice allows us to employ the very same functional form of the standard DF Chandrasekhar formulation, but with the velocity dependence on the perturber velocity in the galactic frame  replaced by the relative velocity with respect to the local medium \citep[see e.g.][for a similar derivation]{Kashlinsky1986}. In addition, with this new prescription, we also discard the tunable $A_{\rm disc}$ parameter previously introduced in \citet{Bonetti2020}. In particular, we obtain

%%%%%%%%%%%%%%%%
\begin{align}\label{eq:adf_disc}
    \mathbf{a}_{\rm df,disc} = -2\pi G^2 \ln(1+\Lambda^2) & m_p \rho_d(R,z) \ \times \nonumber\\
    & \times\left({\rm erf}(X) - \dfrac{2 X{\rm e}^{-X^2}}{\sqrt{\pi}}\right) \dfrac{\mathbf{v}_{\rm rel}}{|\mathbf{v}_{\rm rel}|^3},
\end{align}
%%%%%%%%%%%%%%%%
where $\mathbf{v}_{\rm rel} = \mathbf{v}_p - \mathbf{v}_{\rm rot}(R)$, while $X = |\mathbf{v}_{\rm rel}|/(\sqrt{2}\sigma_R)$, with $\sigma_R$ denoting the radial velocity dispersion. For the Coulomb logarithm, we replace the expression of $p_{\rm min, d}$ appearing in equation~\eqref{eq:pmin_circ} by
%%%%%%%%%%%%%%%%
\begin{equation}\label{eq:pmin_rot}
   p_{\rm min, d} = \dfrac{G m_p}{v_{\rm rel}^2 + \sigma_R^2},
\end{equation}
%%%%%%%%%%%%%%%%
which takes into account both the relative velocity with surrounding medium, that can be different from the circular velocity, and an intrinsic dispersion, which sets the minimum effective distance characterising the encounters between the MP and the objects in the disc.

The presence of a velocity dispersion $\sigma_R$ implies that the disc is not fully rotationally supported. We can characterise the rotational velocity by following \citet{Hernquist1993},

%%%%%%%%%%%%%%%%
\begin{equation}\label{eq:vrot}
    |\mathbf{v}_{\rm rot}|^2 = v_c^2(R) + \sigma_R^2(R) \left(1-\dfrac{\kappa^2}{4\Omega^2} - 2\dfrac{R}{R_d}\right),
\end{equation}
%%%%%%%%%%%%%%%%
in which $v_c$ denotes the circular velocity at $R$, while $\kappa$ and $\Omega$ (both depending on $R$) are the epicyclic and angular frequency, respectively. In evaluating such quantities for multi-component galaxy models, we consider the total potential rather than that generated by the disc only. 
As in \citet{Hernquist1993}, we also assume that the velocity dispersion changes as 
%%%%%%%%%%%%%%%%
\begin{equation}\label{eq:sigma_R}
   \sigma_R^2(R) \propto {\rm e}^{-R/R_d}, 
\end{equation}
%%%%%%%%%%%%%%%%
where the normalisation is evaluated assuming equality between $\sigma_R$ and the critical radial velocity dispersion (augmented by a factor $Q$)\footnote{Here we assume $Q = 1.5$.} at a specific radius, here $2 R_d$, i.e.
%%%%%%%%%%%%%%%%
\begin{equation}\label{eq:sigma_norm}
   \sigma_{R, \rm crit} = Q \times \dfrac{3.36 G \ \Sigma(2 R_d)}{\kappa(2 R_d)}.
\end{equation}
%%%%%%%%%%%%%%%%

We can note that since in equation~\eqref{eq:vrot} the term in parenthesis on the right hand side is usually negative, then the rotational velocity results (often only slightly) lower than the local circular velocity. However, as already pointed out in \citet{Hernquist1993}, at small radii the approximations yielding equation~\eqref{eq:vrot} can break down and provide nonphysical $v_{\rm rot}$ (e.g. an imaginary velocity). To avoid unnecessary complications we decided to fix the rotational velocity equal to a fraction of the local circular velocity (specifically $0.95 v_c(R)$) anytime equation~\eqref{eq:vrot} gives nonphysical results. We expect our choice not to dramatically impact the evolution, since at small radii the dynamics is likely dominated by the stellar bulge (and the associated DF), while the contribution of the disc should be subdominant. We anyway try to address this issues by also directly extracting the rotational profile from an N-body realisation, as we will further discuss in Section~\ref{sec:results}. 

%%%%%%%%%%%%%%%%%%
\begin{figure*}
    \centering
    \includegraphics[width=0.48\textwidth]{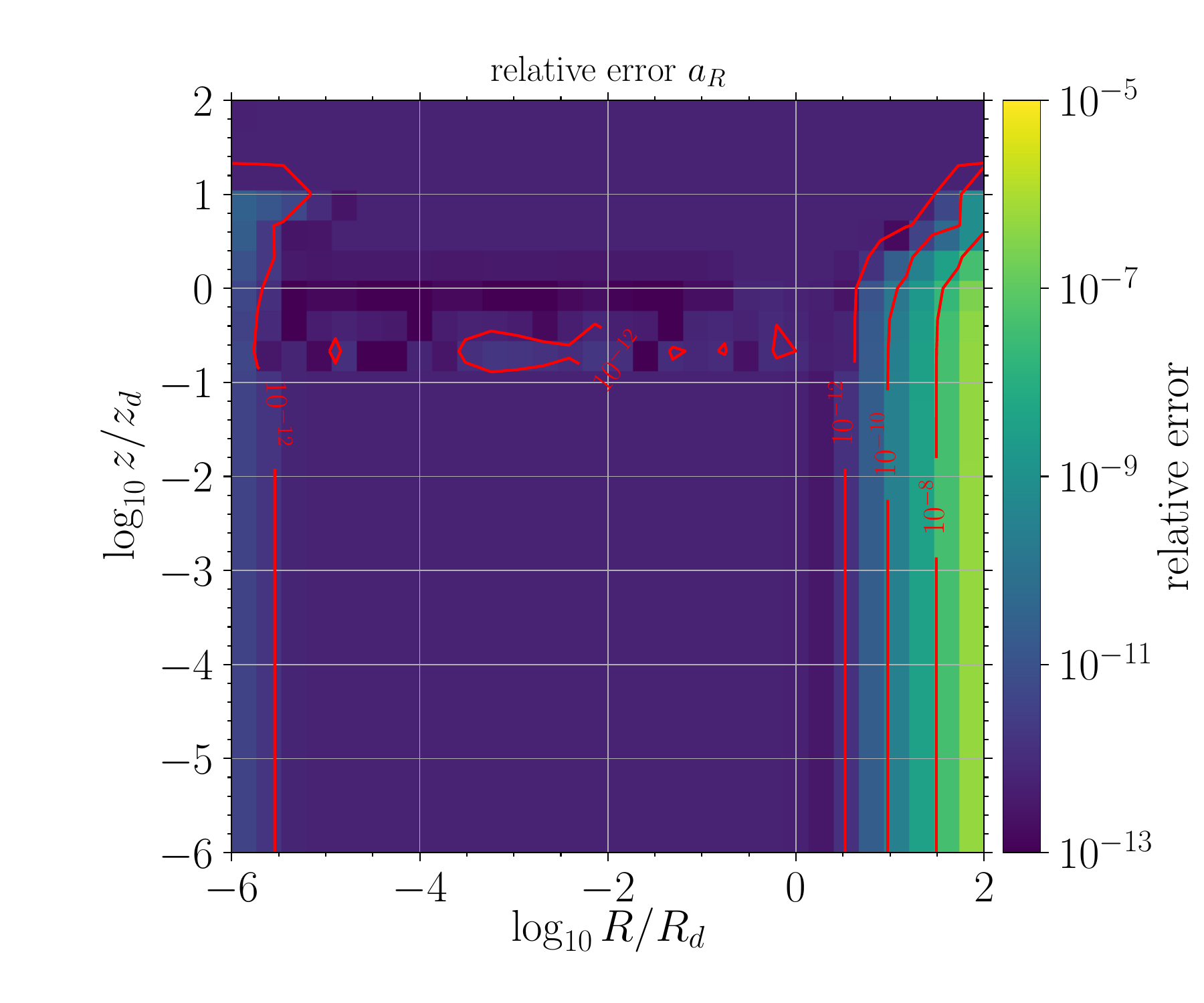}
    \includegraphics[width=0.48\textwidth]{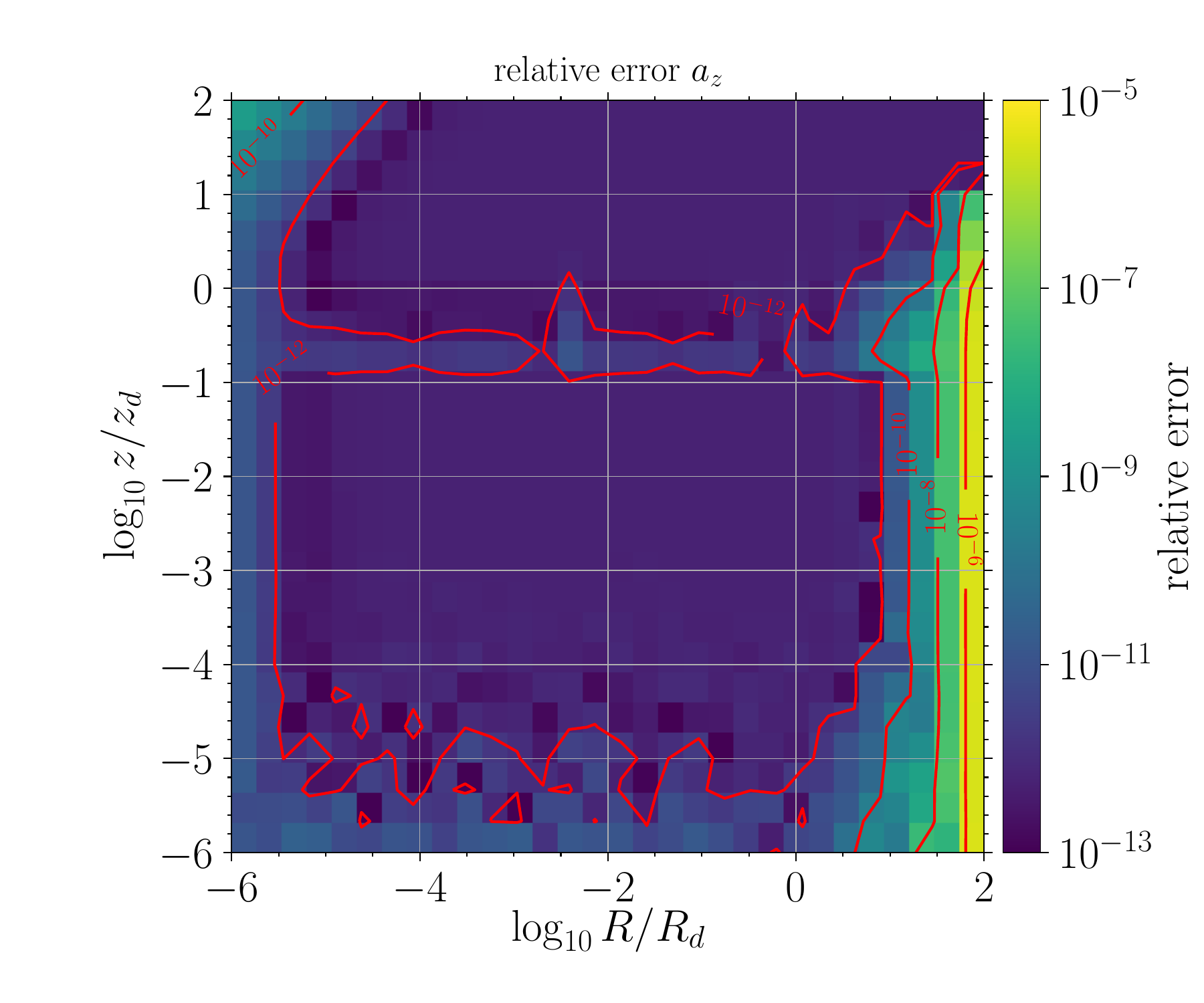}
    \includegraphics[width=0.48\textwidth]{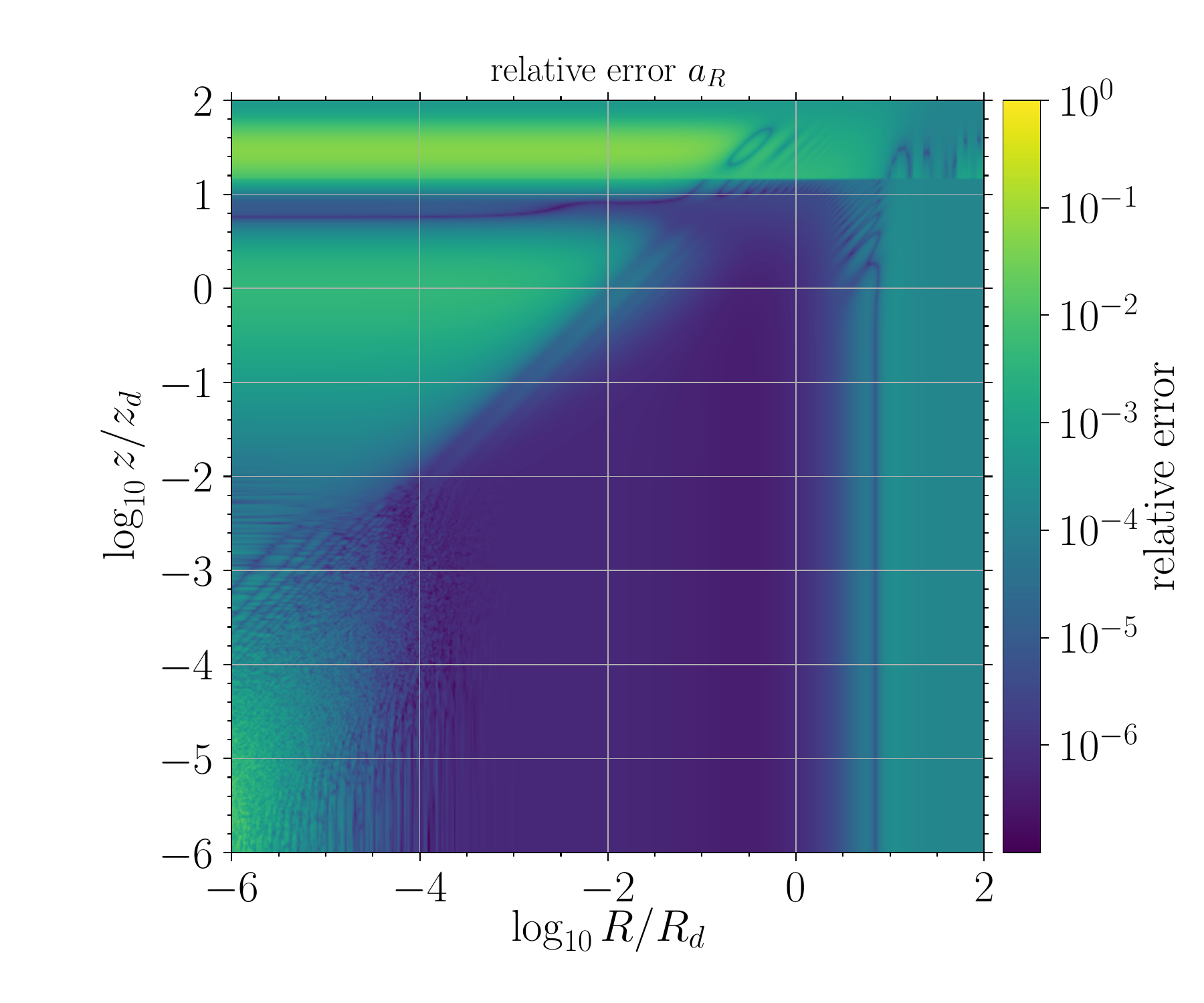}
    \includegraphics[width=0.48\textwidth]{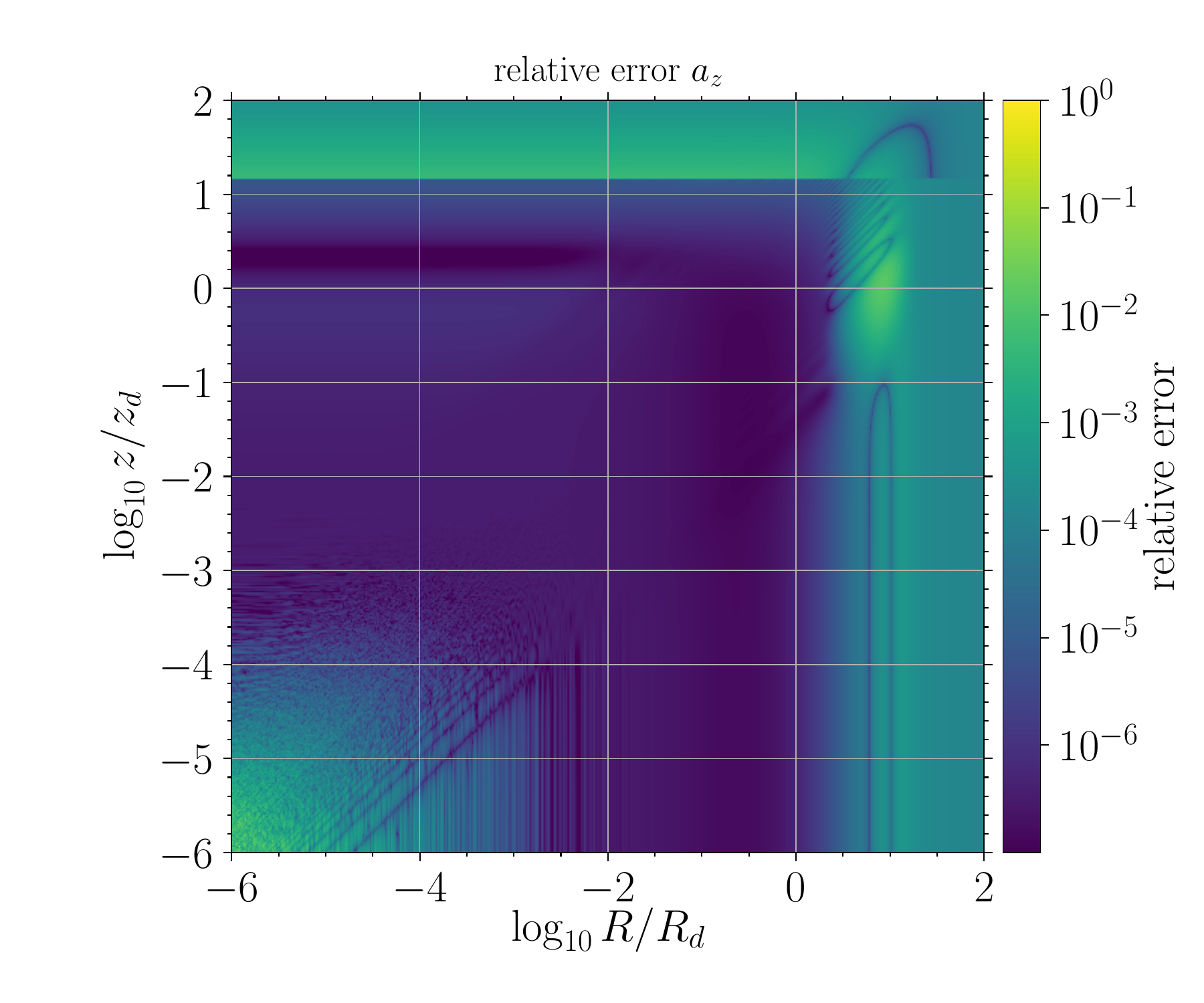}
    \caption{Relative error for the radial (left column) and vertical (right column) accelerations obtained with our code and considering an exponential disc with vertical sech$^2$ profile. Upper panels: relative errors are evaluated against a numerical integration of equations~\eqref{eq:acc_R} and \eqref{eq:acc_z} with $2\times 10^7$ points using trapezoidal rule in the range $k \in [0,10^5]$. Lower panels: relative errors are obtained comparing our acceleration with those computed by the software AGAMA employing an exponential disc with identical parameters.}
    \label{fig:rel_error}
\end{figure*}
%%%%%%%%%%%%%%%%%%

%%%%%%%%%%%%%%%%%%%%%%%%%%%%%%%%%%%%%%%%%%%%%%%%%%%%%%%%%%%%%%%%%%%%%%%%%%%%%%%%%%%%%
\section{Results}
\label{sec:results}

%%%%%%%%%%%%%%%%%
\begin{table}
    \centering
    \begin{tabular}{cccc}
        \hline\hline
        & Halo & Bulge & Disc \\
        \hline\\
        Mass & $1.1 \times 10^{12}\msun$ & $2.2\times 10^{9}\msun$ & $4.4\times 10^{10}\msun$ \\
        Scale Radius & $37 \ (21)$~kpc & $0.96$~kpc & $4.25$~kpc \\
        Vertical scale & -        & - & $0.85$~kpc \\
        Profile & Hernquist (NFW) & Hernquist & Exponential\\ 
        $N_{\rm part}$ & $10^6$ & $5\times 10^5$ & $2\times 10^6$\\
        $\varepsilon$ & 40~pc & 10~pc & 10~pc\\
        \hline\hline
    \end{tabular}
    \caption{Structural parameters of the employed galaxy model. For each component, $N_{\rm part}$ and $\varepsilon$ correspond to the number of particles and the Plummer-equivalent gravitational softening employed in the N-body runs.
    }
    \label{tab:parameters}
\end{table}
%%%%%%%%%%%%%%%%%

In this section, we compare our semi-analytical setup with other similar implementations for the disc potential and the DF treatment, and against full N-body simulations. We start by considering the conservative dynamics in exponential discs, highlighting how a different vertical disc profile can affect the orbits. 
We then analyse our DF prescriptions in multi-component galaxy models and we compare them against the results from N-body simulations featuring the same physical system and MP orbital initial conditions. We employ the public code GIZMO \citep{gizmo} to run the N-body simulations, using the same N-body setup of \citet{Bonetti2020}, whose parameters are reported in Table~\ref{tab:parameters} for completeness.
\footnote{In the framework of N-body simulations of isolated galaxies, the Hernquist profile is generally used to model the dark matter halo instead of the well known NFW profile. This choice is barely related to the finite mass of the Hernquist profile, which instead diverges for the NFW profile. We note that well within their scale radii, the NFW and Hernquist profiles have the very same shape and therefore when focusing on the dynamics inside a galaxy, the two profiles are practically the same \citep{Springel2005}.}

%%%%%%%%%%%%%%%%%%%%%%%%%%%%%%%%%%%%%%%%%
\subsection{Orbits in isolated exponential discs}

Here we investigate the conservative dynamics in single-component galaxies modelled with exponential discs. In our framework we consider both the exponential disc with $z$-profile decaying as ${\rm sech}^2(z/z_d)$ (see Section~\ref{sec:exp_disc}) as well as the so-called double exponential disc, i.e. a profile also decaying as an exponential along $z$.

%%%%%%%%%%%%%%%%%%%%%%%%%%%%%%%%%%%%%%%%%%%%%%%%
\subsubsection{Consistency checks}

%%%%%%%%%%%%%%%%%%
\begin{figure}
    \centering
    \includegraphics[width=0.48\textwidth]{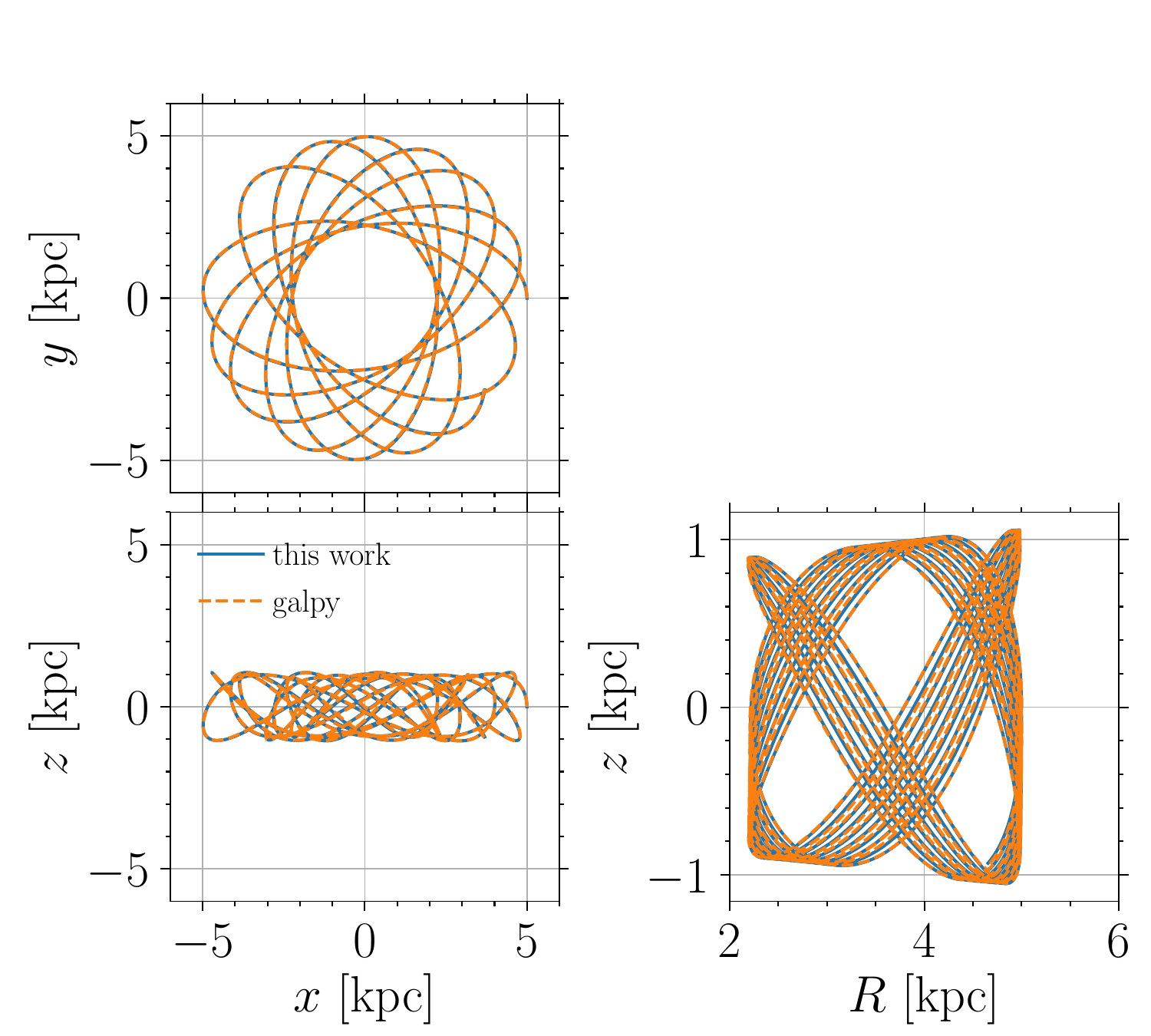}
    \includegraphics[width=0.48\textwidth]{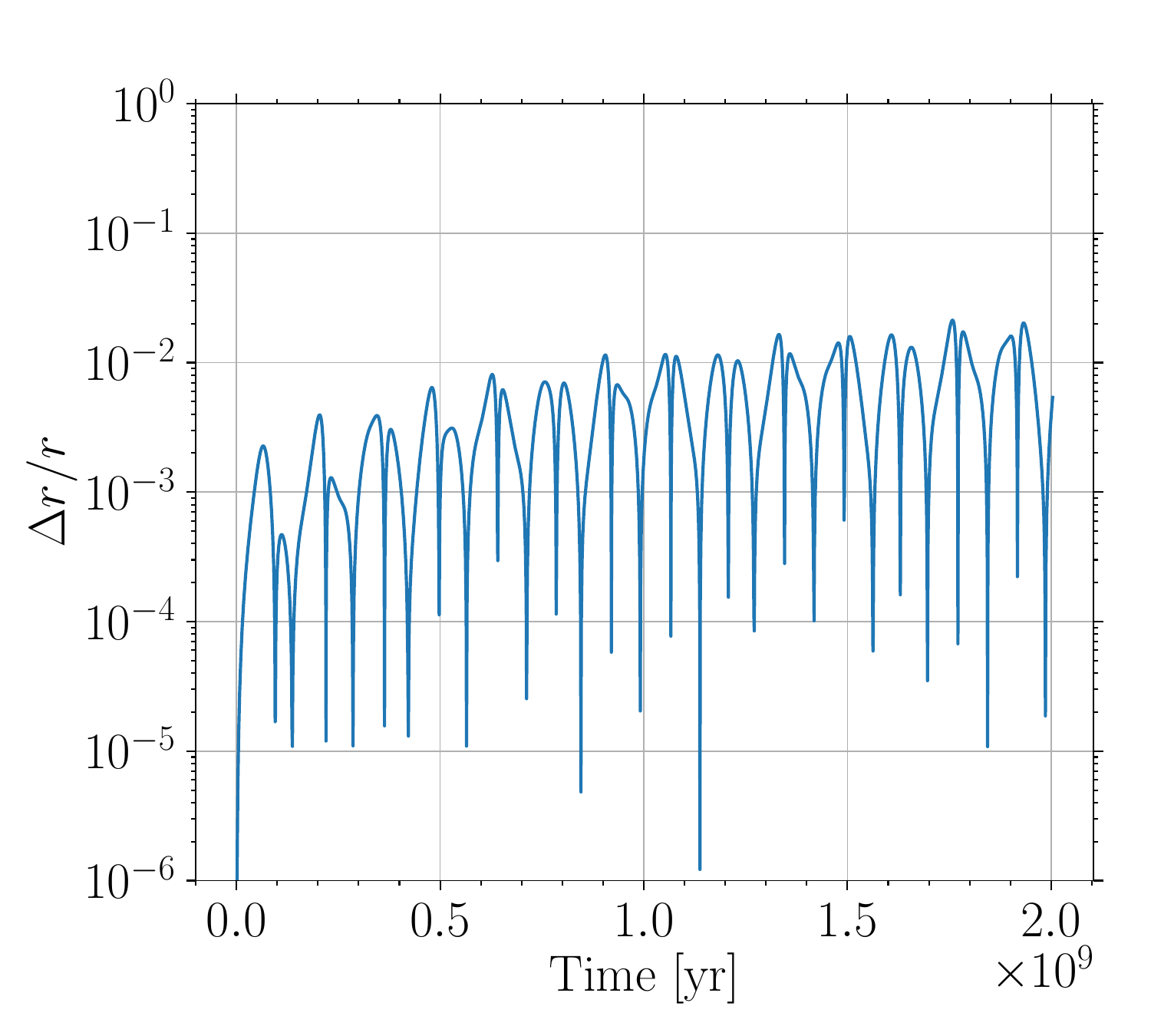}
    \caption{Top: orbit projections in the $x-y$, $x-z$ and $R-z$ planes when considering a ${\rm sech^2}$ exponential disc with mass $M_d = 4.4 \times 10^{10} \msun$, scale radius $R_d = 4.25$ kpc and  $z_d = 0.85$ kpc. We show the comparison between the trajectories obtained within our framework (solid blue lines) and those obtained with the software \galpy~(dashed orange lines). The initial velocity is set equal to  $0.5 v_c$ in both $y$ and $z$ directions, with $v_c$ the circular velocity at a cylindrical radius $R=5$ kpc. 
    Bottom: relative error on the spherical radius between our evolution and that obtained with \galpy. Note how the relative error remains at most at the level of $10^{-2}$ after 2 Gyr of evolution.}
    \label{fig:orbit_disc_galpy}
\end{figure}
%%%%%%%%%%%%%%%%%%

In Fig.~\ref{fig:rel_error}, we report the relative error in the radial (left column) and vertical (right column) accelerations of an exponential disc with a ${\rm sech}^2(z/z_d)$ profile, computed by our semi-analytical framework through a modified version of the double exponential quadrature technique \citep[see e.g.][]{Krzysztof2016}. In the upper panels, we compare our calculation to the numerical evaluation of equations~\eqref{eq:acc_R} and \eqref{eq:acc_z} obtained via the trapezoidal rule with $\sim~2 \times 10^7$ points and assuming a $k_{\rm max} \approx 10^5$. We can appreciate that the relative errors are always very small ($10^{-12}$--$10^{-11}$) over several decades in both $R$ and $z$. Only well above $10 R/R_d$ the relative error starts to grow significantly, but still remaining below $10^{-6}$. This trend is associated with the behaviour of the Bessel function in the integrand function, that oscillates significantly for large $R$, mildly reducing the accuracy of the numerical integration.

In the lower panels, instead, we evaluate the relative errors against the accelerations obtained from the software AGAMA \citep{Vasiliev18}, in which the disc density profile is expanded in multipoles\footnote{For the comparison in Fig.~\ref{fig:rel_error} we adopt within AGAMA a multipole expansion ranging between $10^{-6}R_d$ and 500$R_d$; we used a logarithmically spaced grid in $R$ with 1000 grid points and maximum order of the expansion $l_{\rm max}=100$, while the properties of the $z$ grid are kept as default in AGAMA \citep[see][for more details]{Vasiliev18}.} and the potential (and the corresponding force) is computed via the Poisson's equation for each term of the multipole expansion.
In this case, we find a larger error, but still at the level of $10^{-6}$ for a vast portion of the $(R-z)$ plane, while larger deviations arise at the borders of the plane. Given that the computation in AGAMA relies on the multipole expansion and makes systematic use of spline interpolation, the quantities that AGAMA evaluates are necessarily approximated and subjected to some numerical noise. 
Still, errors are quite small in the relevant portion of the $(R-z)$ plane and reach $10^{-2}$ only for $R$ and $z$ that are much larger or much smaller than $R_d$ and $z_d$. This does not represent a real problem as in realistic galaxy models other galactic components are expected to dominate the dynamics at large/small distances, for instance dark matter halos (at large distances) or stellar bulges (close to the center), hence we can consider our implementation quite robust.

%%%%%%%%%%%%%%%%%%%%%%%%%%%%%%%%%%%%%%%%%%%%%%%%
\subsubsection{Orbit integration in exponential discs}

%%%%%%%%%%%%%%%%%%
\begin{figure}
    \centering
    \includegraphics[width=0.48\textwidth]{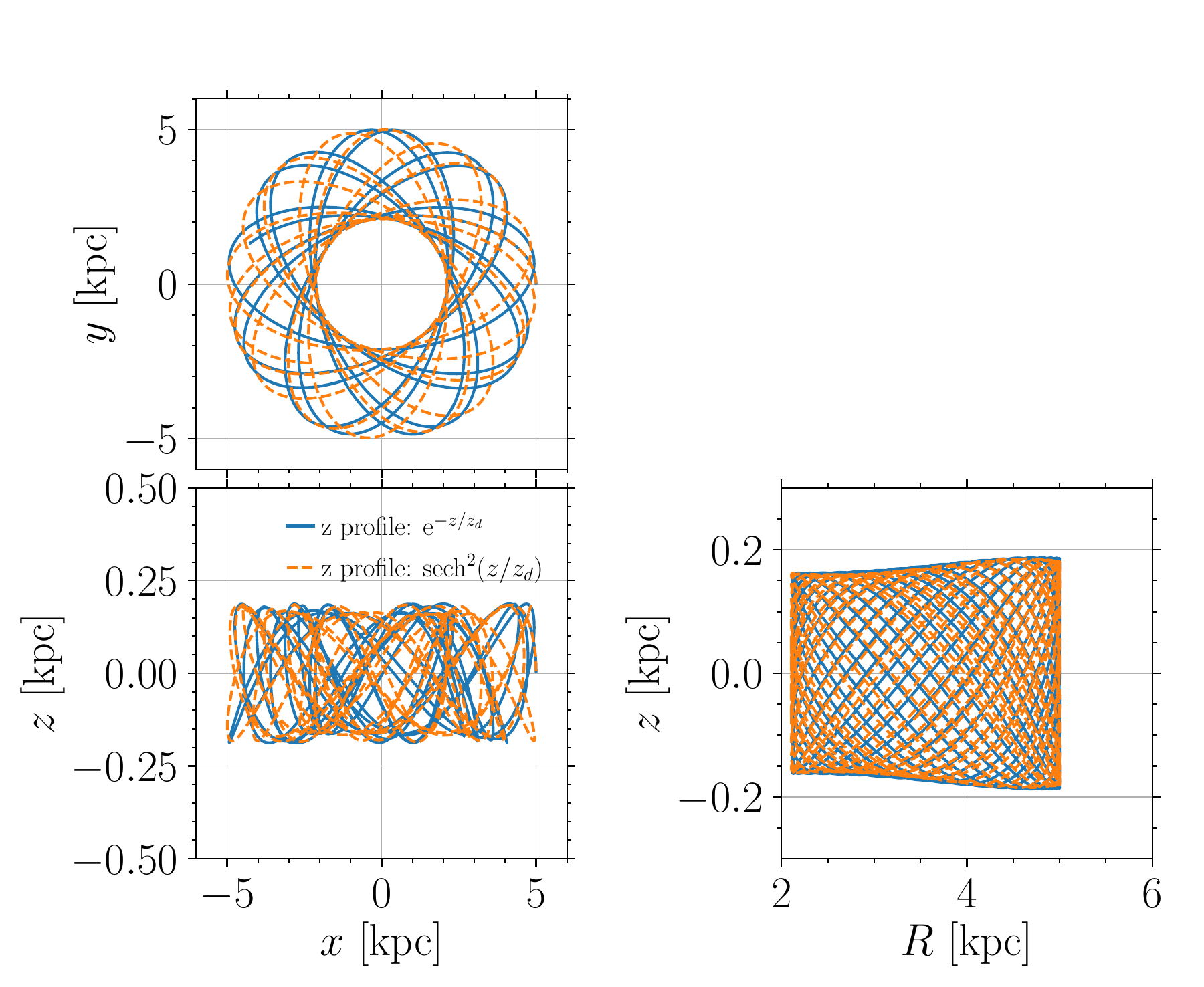}
    \includegraphics[width=0.48\textwidth]{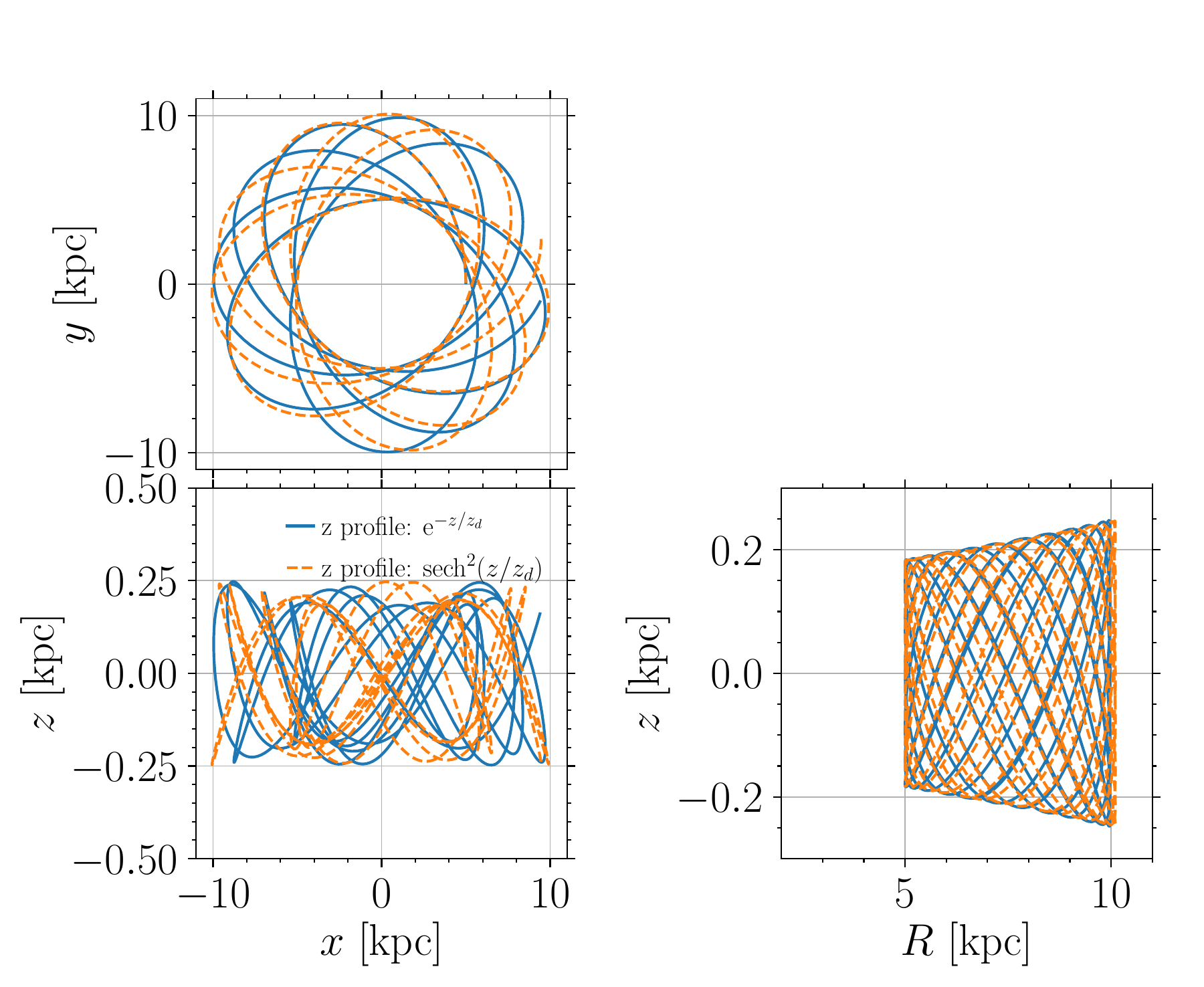}
    \caption{Comparison between orbits in a double exponential disc (solid blue line) and in an exponential disc with a $z$-profile described by $\rm sech^2$ (dashed orange lines). In the top (bottom) panel the in-plane initial velocity is set equal to 0.5 (1.5) the local circular velocity, while vertical velocity is set to 0.1 this value. Each sub-panel shows the orbit projections in the $x-y, x-z$, and $R-z$ planes.}
    \label{fig:orbit_disc}
\end{figure}
%%%%%%%%%%%%%%%%%%

We next compare the orbital evolution that we obtain\footnote{See \citet{Bonetti2016} for details about the orbit integration method.} with those of the software \galpy  \ \citep{galpy}. 
Specifically, we analyse the trajectories obtained in an exponential disc with mass $M_d = 4.4 \times 10^{10} \msun$, scale radius $R_d = 4.25$ kpc and $z_d = 0.85$ kpc and with a vertical profile given by either ${\rm sech}^2(z/z_d)$ or ${\rm e}^{-|z|/z_d}$ (see also Tab.~\ref{tab:parameters}).
In the top panel of Fig.~\ref{fig:orbit_disc_galpy}, we compare the orbit for the ${\rm sech^2}$ case (blue solid for our implementation vs orange dashed for \galpy)\footnote{The ${\rm sech^2}$ exponential disc within the \galpy~framework is accessible through the SCF basis-function-expansion (see \galpy~documentation).} by showing the trajectory projections in the $x-y$, $x-z$, $R-z$ planes. The orbit is evolved for approximately 2 Gyr starting from an initial radial separation of $R = 5$ kpc and an initial velocity along the $y$ and $z$ directions equal to 0.5 times the local circular velocity. From the top panel, we can infer that the orbit in our implementation closely matches that of \galpy~ and to better quantify the agreement, in the bottom panel of the figure, we evaluate the relative error on the spherical radius between the two runs, finding that is at the level of $10^{-2}$ at most. This small difference is probably due to the different approaches followed to compute the disc potential: specifically through the employment of the Hankel's transform in our case, which reduces the problem to the numerical computation of an integral (see equation~\ref{eq:disk_pot}), or via a multipole expansion approximation in the \galpy~framework. We perform the same comparison by also considering the double exponential disc profile. As for the ${\rm sech^2}$ we found a good agreement, therefore enforcing the robustness of our framework.
Another important comparison with \galpy~concerns the computational cost, that for a semi-analytical framework has to necessarily be modest in order to fully exploit its power.
For each run, we find a total run-time of the order of few tens of seconds (on a single core of a standard laptop), similar to that of \galpy~ (but only once an adequate accuracy, similar to that in our framework, is also adopted in \galpy). Moreover, to further speed-up the orbital integration, we have also implemented the possibility of storing in advance the accelerations computed on a $(R-z)$ adaptive grid and then employ a bi-cubic interpolation to obtain $a_R$ and $a_z$ on arbitrary points. 

Finally, we compare the orbital shape of a point mass when subject to the potential of a double exponential disc and a ${\rm sech}^2$ one. We consider the same disc and initial conditions as described above, but here we set the velocity in the $z$ direction equal to 0.1 times the local circular velocity. The results are shown in Fig.~\ref{fig:orbit_disc}. In both the left and right panels we show the three projections over the $x-y$, $x-z$ and $y-z$ planes as sub-panels (from top left to bottom right). We can appreciate that the orbits resemble each other, at least qualitatively, in the sense that they cover more or less the same volume of the disc. Still, from a quantitative point of view, e.g. checking the position reached at a specific time, the orbits are well distinct and evolve on slightly different timescales. Hence, a precise reconstruction of orbits in a realistic galaxy requires the employment of the density/potential pair that better describes such structure, that, given the variety of profiles observed, would require an ad-hoc model for each individual case. Such exercise can be easily preformed using our procedure, that provides both double exponential and ${\rm sech}^2$ exponential disc implementations, but is beyond the scope of our current validation study.

%%%%%%%%%%%%%%%%%%%%%%%%%%%%%%%%%%%%%%%%%
\subsection{DF in multi-components models}

%%%%%%%%%%%%%%%%
\begin{figure}
    \centering
    \includegraphics[width=0.48\textwidth]{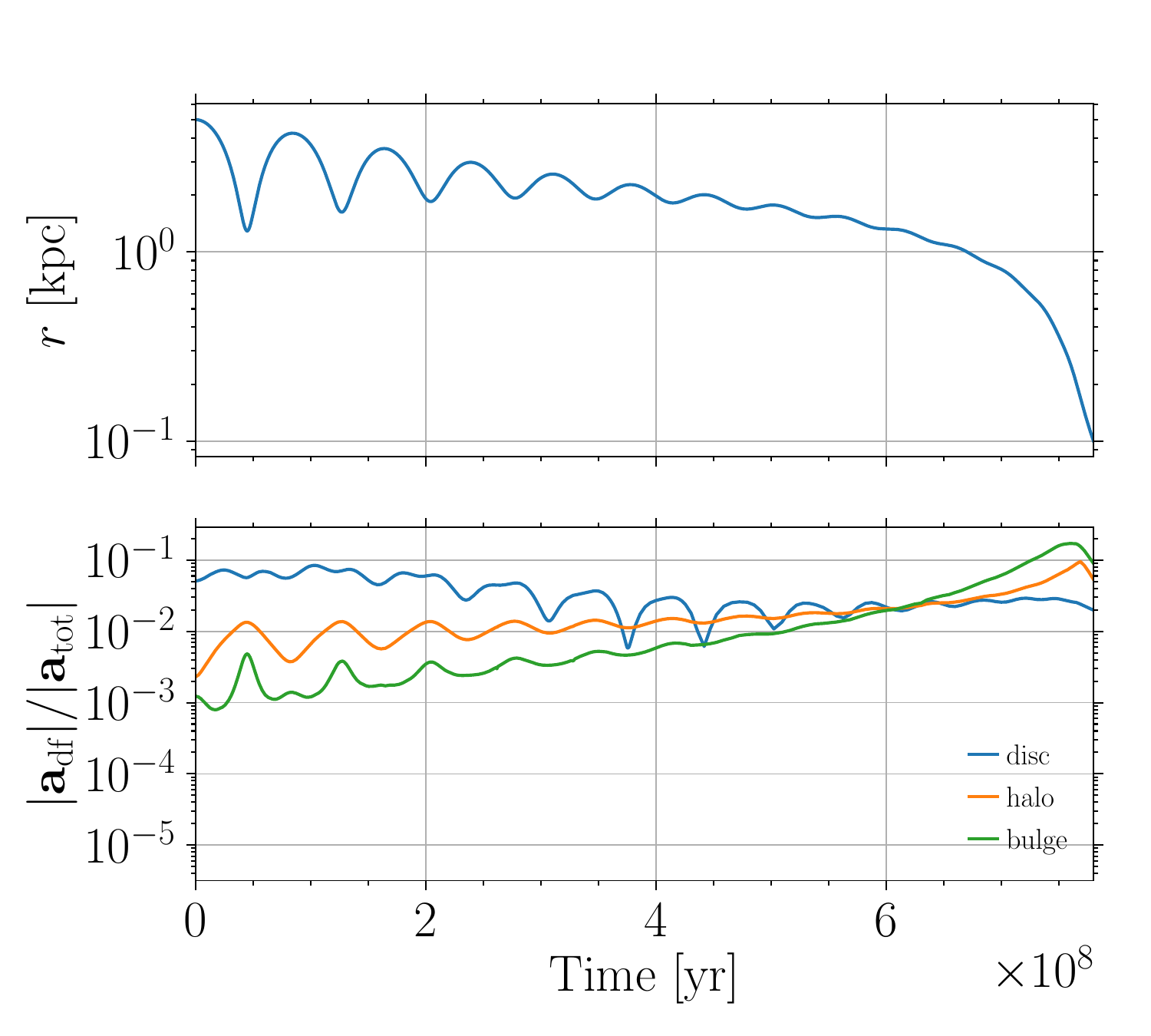}
    \caption{Upper panel: MP radial evolution. Lower panel: relative contributions of the DF accelerations from the three galactic components (each acceleration modulus is normalised to the total acceleration, i.e. both dissipative and conservative).}
    \label{fig:DF_contr}
\end{figure}
%%%%%%%%%%%%%%%%

%%%%%%%%%%%%%%%%%%
\begin{figure*}
    \centering
    \includegraphics[width=0.48\textwidth]{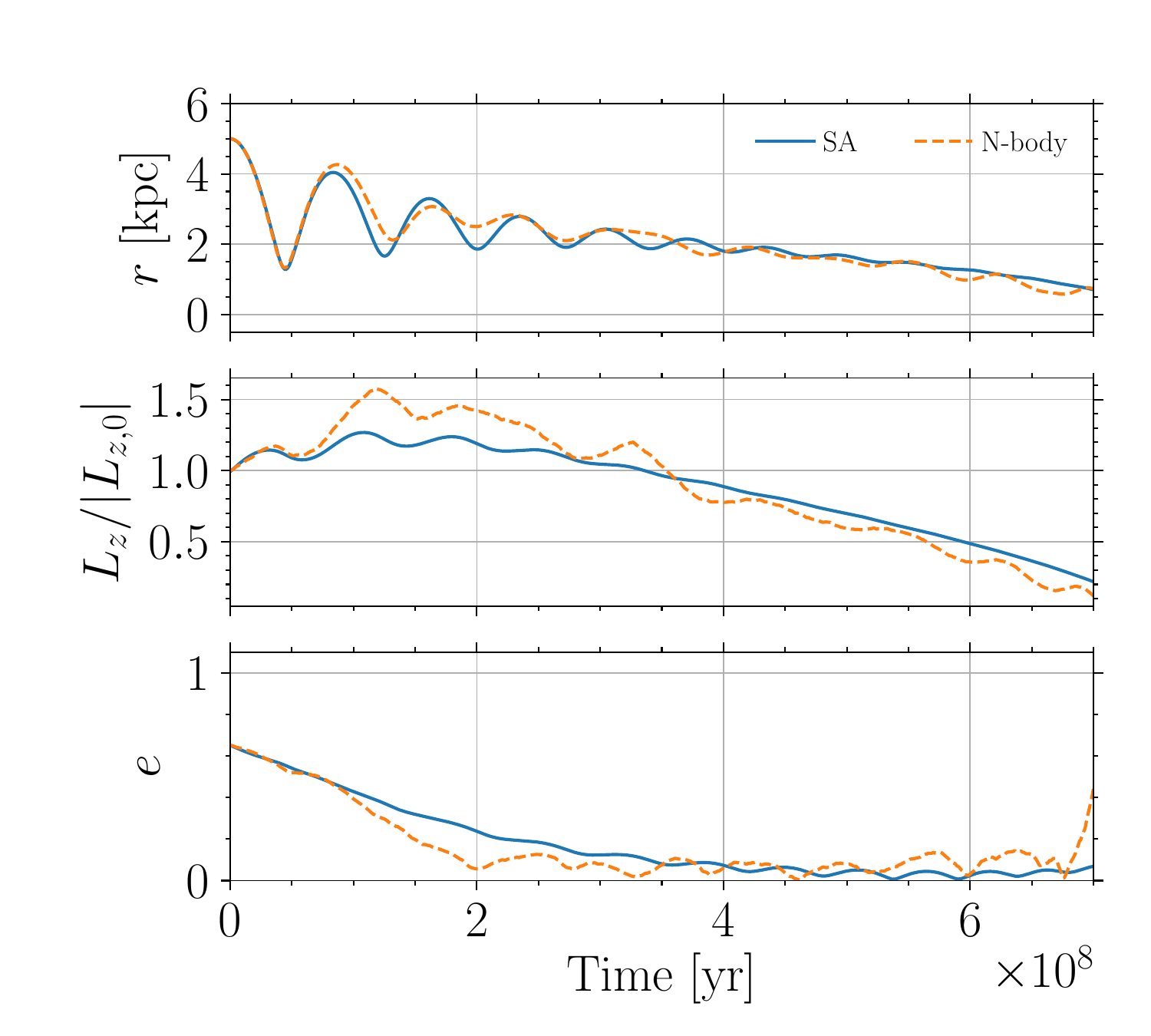}
    \includegraphics[width=0.48\textwidth]{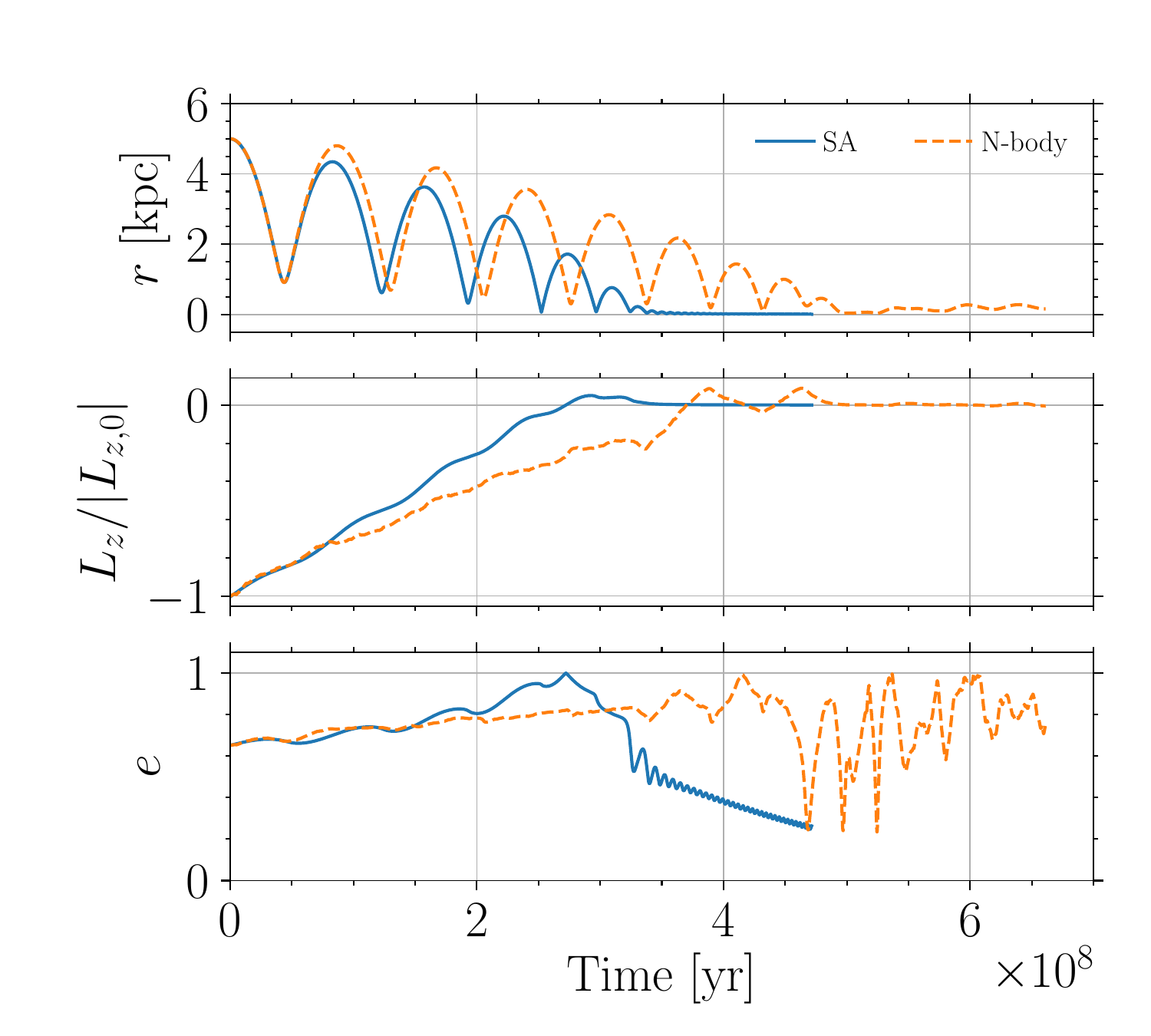}
    \caption{Time evolution of the radial separation, the normalised angular momentum, and the eccentricity obtained with the semi-analytical framework (SA, blue solid lines) and an N-body simulation (orange dashed lines), employing the same initial conditions and physical parameters. Left panel: co-planar prograde orbit. Right panel: co-planar initially retrograde orbit.}
    \label{fig:multi_plane_standard}
\end{figure*}
%%%%%%%%%%%%%%%%%%

%%%%%%%%%%%%%%%%%%
\begin{figure*}
    \centering
    \includegraphics[width=0.48\textwidth]{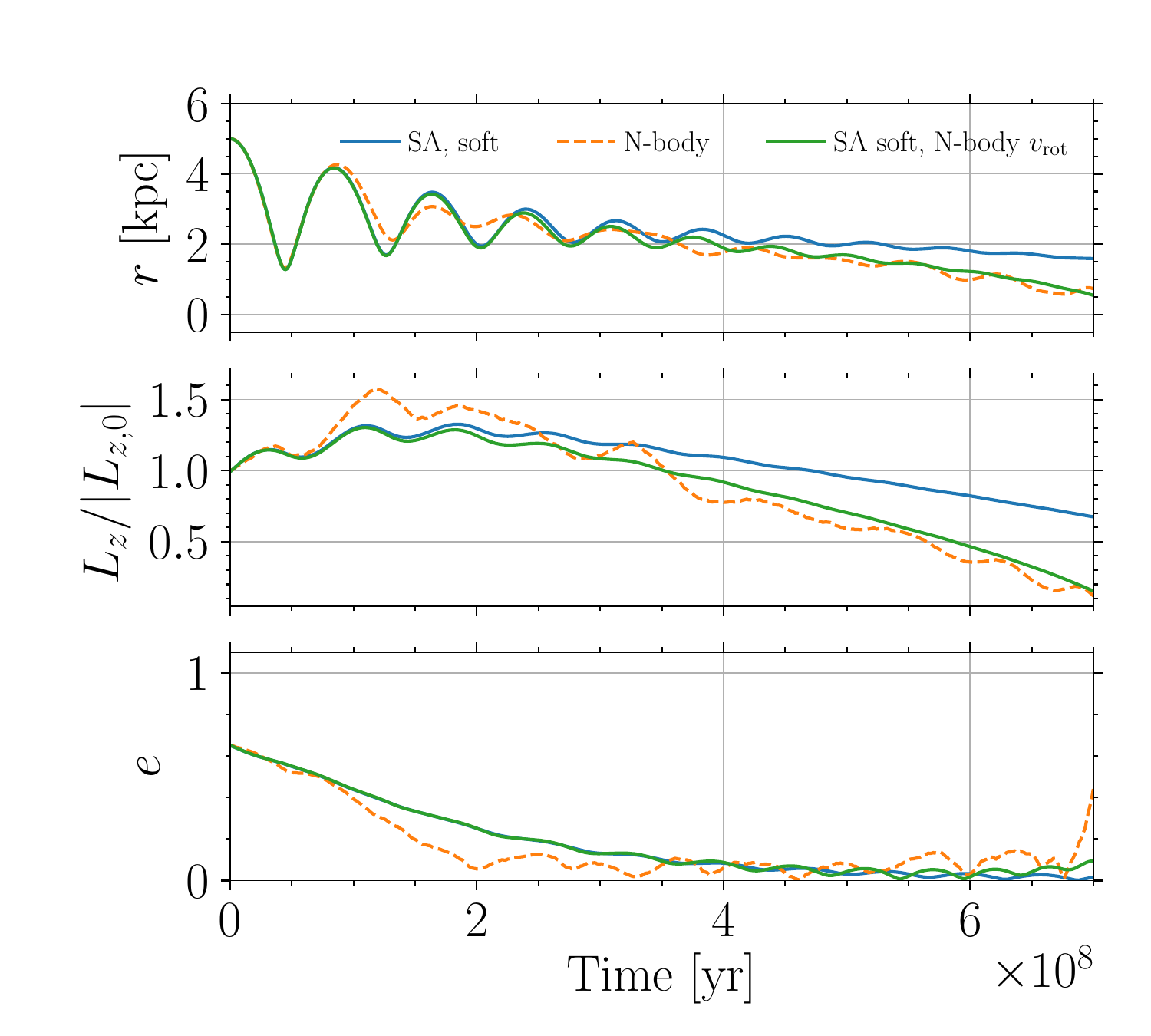}
    \includegraphics[width=0.48\textwidth]{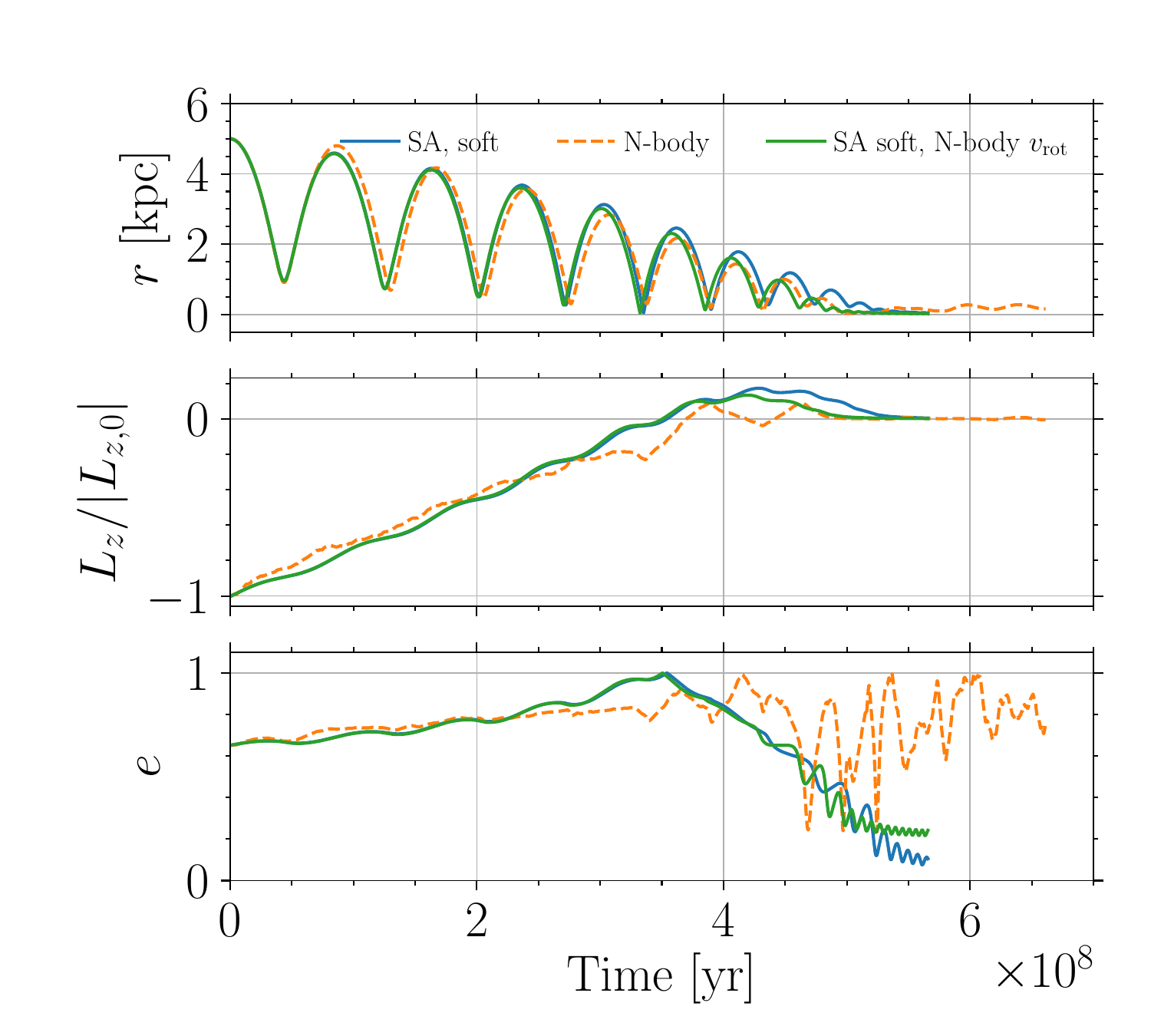}
    \caption{Same as Fig.~\ref{fig:multi_plane_standard}, but modifying the semi-analytical setup to take into account the finite spatial resolution of the N-body simulations (solid blue lines) as well as the rotational velocity in the N-body run (solid green lines). 
    Specifically, the minimum impact parameter $p_{\rm min}$ for each component (disc, halo, bulge) is fixed at 2.8 times the respective Plummer-equivalent softening length, while the evolution described by the green lines also employs the rotational velocity directly extracted from the N-body simulations.
    }
    \label{fig:multi_plane_soft}
\end{figure*}
%%%%%%%%%%%%%%%%%%

Now, we consider the MP evolution under the DF combined effect of all galactic components. Before comparing our semi-analytical evolution with that of an N-body simulation, we first assess the relative strength of DF from different galactic components, specifically halo, bulge and disc. In the upper panel of Fig.~\ref{fig:DF_contr} we show a generic MP radial decay achievable within our framework, while in the bottom panel we show the corresponding instantaneous relative contribution of DF generated by each of the considered galactic component (see labels) normalised to the total acceleration suffered by the MP. From the bottom panel we can clearly appreciate that, for a galaxy with structural parameters chosen as in Tab.~\ref{tab:parameters}, our framework predicts an early evolution where the dominant DF is actually generated by the disc component, and we therefore have to expect specific phenomenology connected to the net rotation of the disc structure, for instance the ``drag towards circular co-rotation''. Although sub-dominant contributions by the halo and bulge components are also present, only the halo has a non-negligible effect, being the bulge too compact to properly affect the MP orbit quite outside the scale radius. However, as the evolution proceeds, the above picture reverses. As soon as the MP reach a radial separation of $\sim 2$ kpc (comparable with the bulge length-scale), the DF contribution from disc weakens and DF from spherical components takes over, especially the contribution from the stellar bulge that dominates the MP orbital decay at small scales. 
From the above findings, we expect, in a generic disc galaxy modelled with at least three componenents, the evolution of a MP to be initially driven by the interaction with the disc, therefore making its modelisation quite relevant.

We now proceed to validate our semi-analytical implementation against full N-body simulations.
In Fig.~\ref{fig:multi_plane_standard}, we show, from top to bottom, the time evolution of radial separation, the $z$ component of the angular momentum (normalised to its initial value), and the eccentricity of a MP orbiting in the galactic mid-plane (i.e. co-planar with the galactic disc). \footnote{As in \citet{Bonetti2020}, the eccentricity is found by assuming instantaneous conservation of energy and angular momentum and solving for the two radial roots (i.e. peri- and apocentre) such that the radial velocity vanishes. The orbital inclination is instead computed by finding the angle between the angular momentum and the $z$ axis.}
We here compare the evolution obtained with our semi-analytical setup, as described in Section~\ref{sec:method} (solid lines), and with N-body simulations (dashed lines). 
From the figure, we can infer that our semi-analytical setup can qualitatively catch the relevant phenomenology of the evolution displayed by the N-body simulations, such as the trends shown by the angular momentum and eccentricity. 
For instance, in the prograde case we clearly observe that DF is responsible for a quite rapid circularisation of the MP orbit well before the MP enters the bulge-dominated region ($\lesssim 1$ kpc), while in the initially retrograde case, first DF greatly increases the orbital eccentricity until the angular momentum reverses, then when the orbit becomes prograde it again promotes the circularisation. Despite we can capture the behaviour of the N-body runs, still the timescales are not exactly reproduced. 
Indeed, although the evolution of a co-planar prograde orbit (left panel) seems to be well recovered by our framework, the same does not hold when looking at an initially co-planar retrograde trajectory (right panel), in which the semi-analytical decay looks faster.
Such differences, however, can be interpreted by reminding that also the N-body simulations have some limitations and a fair comparison with our semi-analytical framework has to necessarily take into account the limited spatial and mass resolution of N-body runs. Specifically, we need to consider that in N-body codes all interactions below the softening length get considerably weakened and, as a consequence, the resulting DF force acting on a MP is smaller with respect to an ideal and arbitrarily resolved physical system \citep[see, e.g.][]{Tremmel2015,Pfister2017}. 

%%%%%%%%%%%%%%%%%%
\begin{figure}
    \centering
    \includegraphics[width=0.48\textwidth]{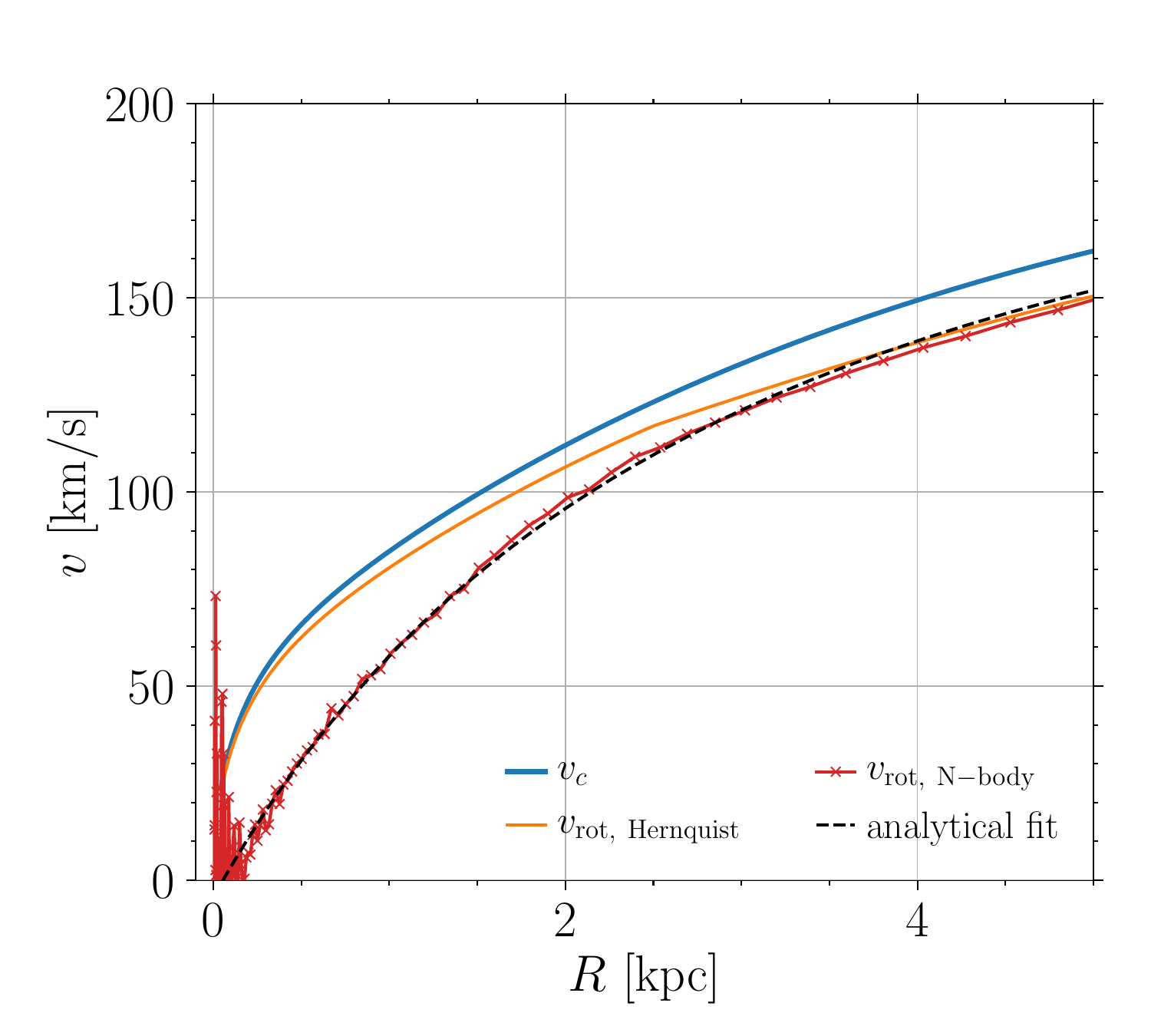}
    \caption{Circular velocity (thick blue line), rotational velocity as predicted by equation~\eqref{eq:vrot} (thin orange line) and extracted from N-body simulation (red crossed line) as a function of radius. At small $R$, the large oscillations of N-body $v_{\rm rot}$ are due to the limited particle coverage in the central regions. To avoid spurious numerical noise we employed an analytical fit of the N-body $v_{\rm rot}$ (black dashed line).}
    \label{fig:Vrot}
\end{figure}
%%%%%%%%%%%%%%%%%%

%%%%%%%%%%%%%%%%%%
\begin{figure*}
    \centering
    \includegraphics[width=0.48\textwidth]{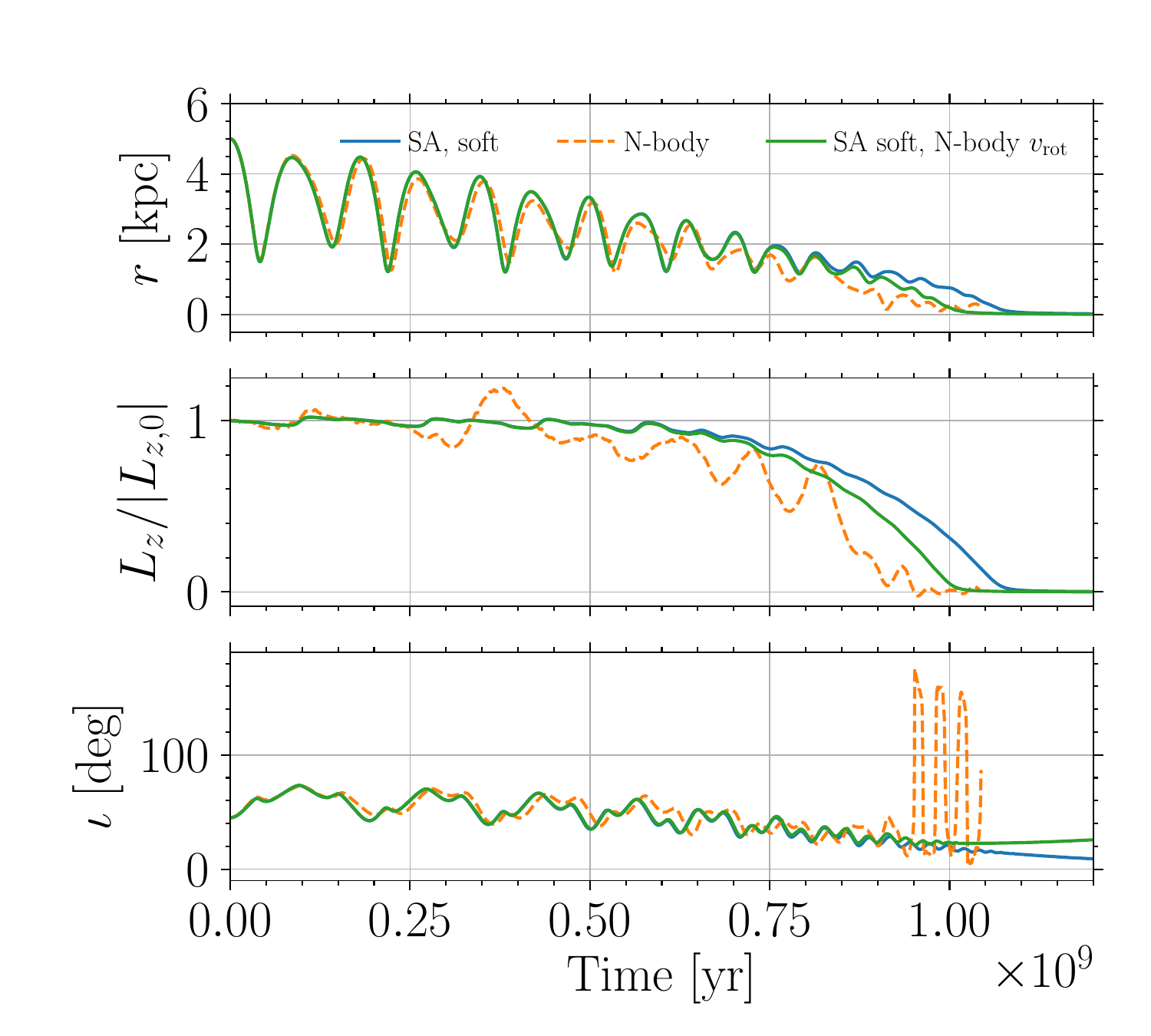}
    \includegraphics[width=0.48\textwidth]{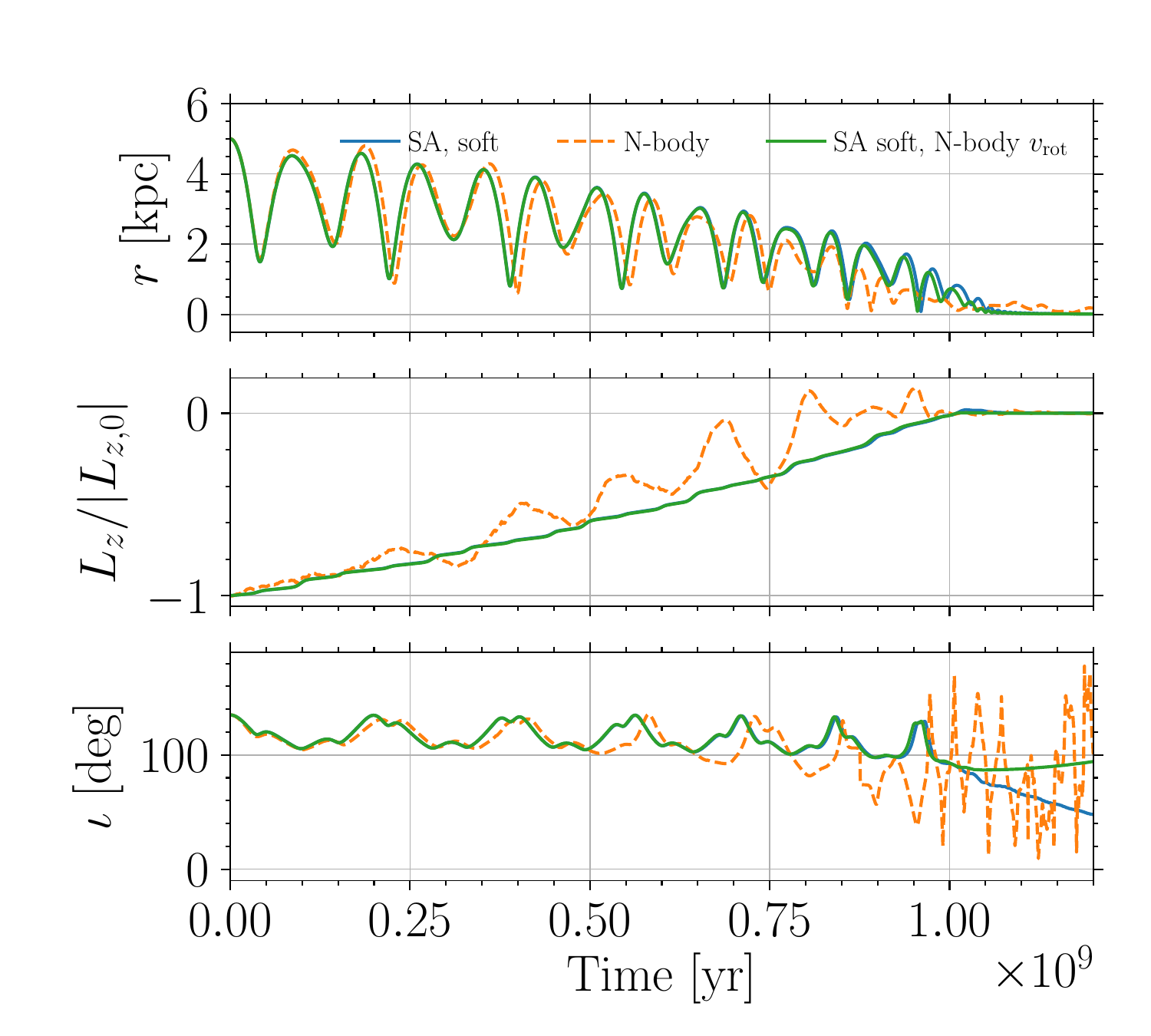}
    \includegraphics[width=0.48\textwidth]{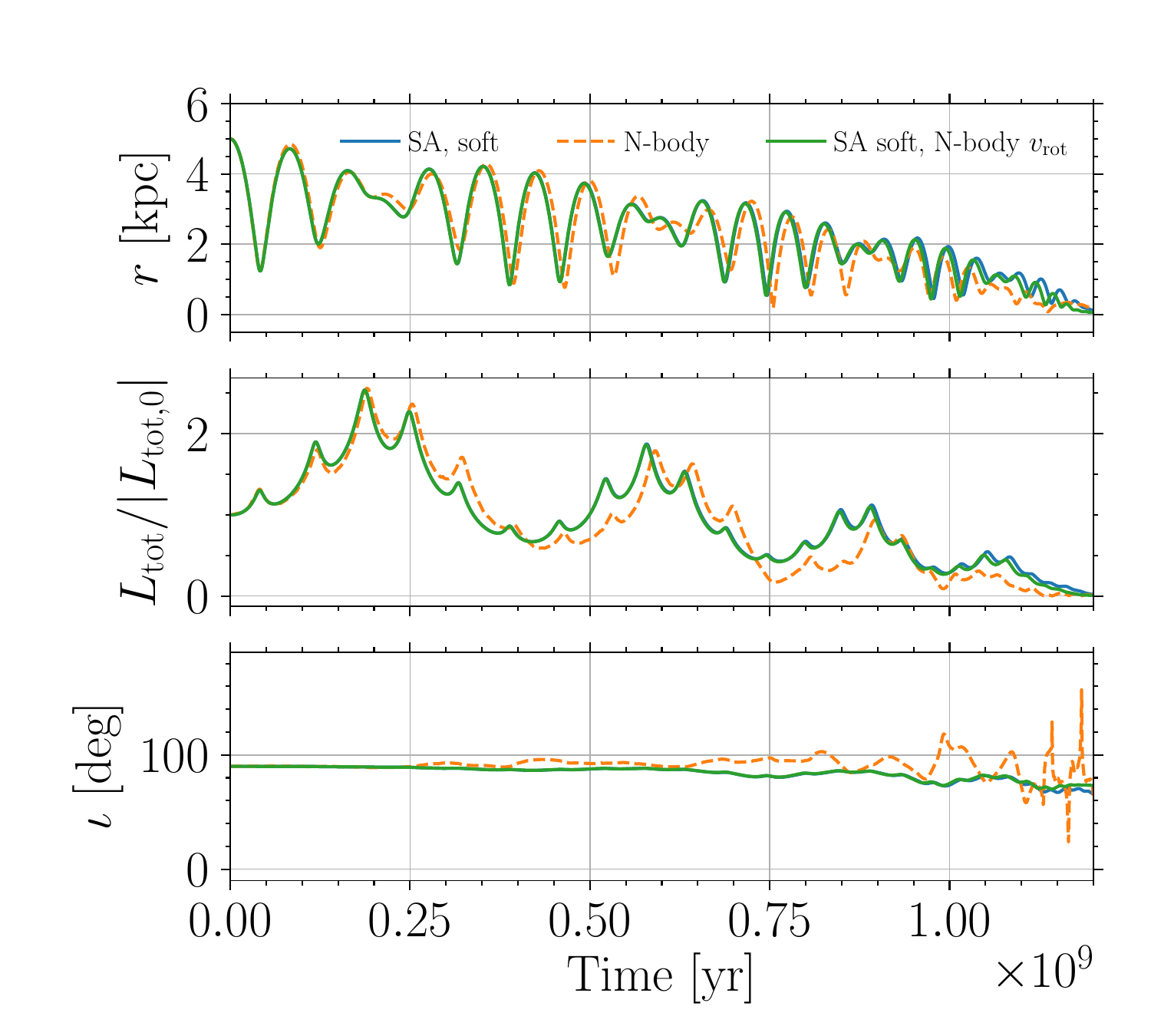}
    \caption{Same as Fig.~\ref{fig:multi_plane_soft}, but considering three different initial inclinations, $45^\circ$ (upper left), $135^\circ$ (upper right), $90^\circ$ (bottom). The bottom panel of each case shows the inclination evolution, while for the $\iota = 90^\circ$ run the total angular momentum is shown instead of its $z$ component.}
    \label{fig:multi_inc}
\end{figure*}
%%%%%%%%%%%%%%%%%%

To account for the limited N-body resolution we can act on the parameters that enter the expression of the DF force. According to equation~\eqref{eq:pmin_rot}, we can observe that the minimum impact parameter $p_{\rm min}$ depends on the MP velocity and specifically for the disc DF it depends on the relative velocity between the MP and the surrounding medium. A small $p_{\rm min}$ translates into larger DF accelerations  through the term $\ln(1+\Lambda^2)$. For the retrograde case, $v_{\rm rel}$ can be quite large, therefore $p_{\rm min}$ can decrease significantly, becoming smaller than the softening length of the N-body simulation. To assess the influence of the softening in the N-body framework we thus force the minimum impact parameter $p_{\rm min}$ to be larger than 2.8 times the Plummer-equivalent softening length of the corresponding particle type, i.e. halo, bulge or disc. The evolution resulting from softened semi-analytical setup is shown in Fig.~\ref{fig:multi_plane_soft} with blue lines. Both the prograde and retrograde orbits (left and right panel respectively) now show a slower decay, and in particular the retrograde case appears to be much more similar to the N-body evolution. The prograde inspiral however seems now too slow.

Another important effect that needs to be considered is the possible difference between the rotational velocity profile employed in our semi-analytic approach, based on the approximated equation~\eqref{eq:vrot} \citep{Hernquist1993}, and that of the simulation. 
As shown in Fig.~\ref{fig:Vrot}, a direct comparison of the two velocity profiles reveal that, as we move to smaller radii, the difference in the rotational velocity profile between the one used in the semi-analaytical approach and that in the N-body simulation gets enhanced. 
If the rotational velocity is significantly different from the local circular one when the orbit becomes nearly circular (i.e. above 400 Myr in the prograde case), then the disc DF force does not vanish (see equations~\ref{eq:adf_disc} and \ref{eq:vrot}), resulting in a slightly faster decay. In order to account for this effect, we implemented the rotational velocity profile extracted from the simulation and fitted with a rational function of the radius (see the dashed black line in Fig.~\ref{fig:Vrot}) 
%%%%%%%%%%%%%%%%%%
\begin{equation}
    |\mathbf{v}_{\rm rot}| = \dfrac{a (R^2 + b R)}{R^2 + c R + d} + K \quad {\rm [km/s]}
\end{equation}
%%%%%%%%%%%%%%%%%%
with $a = 1.25\times 10^2, b = 1.24\times 10^2, c = 6.03 \times 10^1, d = 1.91\times 10^2, K = -4.41$. This modification is enough to improve the agreement between the the two approaches, as shown by the green lines in Fig.~\ref{fig:multi_plane_soft}. Moreover, the apparent good agreement between the N-body run and the semi-analytical approach in the left panel of Fig.~\ref{fig:multi_plane_standard} actually results from a fortunate coincidence related to the different weighting of the DF acceleration generated by each galactic components, specifically a slightly stronger (because not softened) DF from the spherical components and a weaker DF from the disc, due to a velocity profile closer to the circular one. 

Although the orbits we have considered so far were co-planar with the disc, our semi-analytical framework allows us to explore orbits with arbitrary inclinations (see Section~\ref{sec:exp_disc} for details). Here we check the orbital decay of MPs starting from off-plane configurations, as reported in in Fig.~\ref{fig:multi_inc}, $\iota = 45^\circ$ (upper left panels), $\iota = 135^\circ$ (upper right panels) and $\iota = 90^\circ$ (bottom panels); see additional off-plane cases shown in Appendix~\ref{sec:appD}. For each case, instead of the eccentricity, we show here the evolution of the inclination. For the $\iota = 90^\circ$ case we also report the total angular momentum evolution, being $L_z \approx 0$. As in Fig.~\ref{fig:multi_plane_soft}, for each inclination we compare the evolution derived from N-body simulations (orange dashed lines) with that computed in our semi-analytical framework, again taking care of including the softening of the corresponding N-body run (solid blue lines) and the N-body rotational profile (solid green lines). From the figure, we can clearly see that the semi-analytical evolution matches remarkably well the N-body one, correctly reproducing both the observed phenomenology and the decay timescale. Minor differences are of course present, but these are unavoidable given the discrete nature and the limited mass resolution of the N-body approach. Similarly to the planar counterparts, also in these inclined runs a better agreement is met when both the softening and the rotational curve of the N-body are considered. 

We stress again that the very good agreement that we observe in both planar and non-planar cases is achieved only when some input from the N-body runs is provided. Indeed, a fair comparison has to necessarily take into account the limited resolution of N-body simulations. Still, for realistic galaxies we expect that a faster decay, similar to the one we find when considering our standard framework (see Fig.~\ref{fig:multi_plane_standard}), is actually more physical and should provide a better timescale estimate, valid at least down to radial separations not much smaller than the bulge scale radius. Indeed, well inside this scale radius we expect the correction from fast moving star to start being relevant, hence a more general DF formulation should be used.
Despite this, our framework has proved to excellently capture the physics driving MP motion in realistic multi-component galaxy models with rotationally supported structures. Moreover, given its quite modest computational cost compared to full N-body simulations, our framework represents a particularly attractive solution for large parameters space explorations in terms of galaxy structures as well as MP masses and orbital configurations. 

%%%%%%%%%%%%%%%%%%%%%%%%%%%%%%%%%%%%%%%%%%%%%%%%%%%%%%%%%%%%%%%%%%%%%%%%%%%%%%%%%%%%%
\section{Discussion and conclusions}
\label{sec:discussion}

In this paper, we developed a semi-analytical framework to integrate the motion of particles in realistic galactic potentials composed of multiple components. Among several implemented potential-density pairs (spherical, axis-symmetric and triaxial), we included the cases of exponential discs with vertical density decaying either as ${\rm e}^{-|z|/z_d}$ or ${\rm sech}^2 (z/z_d)$, for which we developed an efficient setup to numerically compute its potential and, consequently, the acceleration field. 
Though more complex to tackle, these exponential profiles have to be preferred to other analytical profiles, that can only partially reproduce the exponential decay of the surface brightness of observed disc galaxies. For example, the Miyamoto-Nagai \citep[MN,][]{Miyamoto1975} disc is one of the most common analytical profiles employed to model the potential of disc galaxies. Still, this profile can significantly differ than the one of realistic galaxies, especially at large radii where it falls off as a power law rather than exponentially \citep[see discussion in chapter 2 of][]{BT}. To enhance the density decay, a combination of three MN discs is usually employed, although this requires one of the discs to have a negative mass, often leading to the appearance of non-physical negative density regions in the $(R-z)$ plane \citep[see e.g.][]{Barros2016,Bonetti2019}.  
This strongly supports our choice about the direct employment of the exponential disc profile.

In addition to a more realistic description of the density and potential of disc galaxies, our semi-analytical framework can accurately reproduce the orbital decay of MPs due to DF. For an arbitrary mass distribution, an analytical description of the MP orbital decay from first principles is generally unfeasible, since such complex phenomenon depends both on local and non-local effects \citep[see, e.g., the recent discussion in][]{Tamfal20}. Any semi-analytical formulation must therefore introduce a number of approximations. We leverage on the fact that generally, at a first level of approximation, local effects are more relevant in the vast majority of common astrophysical situations. This allows us to employ the formalism derived by \citet{Chandra43} based on two-body interactions.
We described the DF for both the spherical profiles and the disc. For the latter, which features a net rotation, we describe DF as a function of the relative velocity between the MP and the local surrounding medium. We compared the evolution in the semi-analytical framework against full N-body simulations, and we showed that, when additional information about the limited resolution of the latter is included in our setup, the agreement between the two is excellent.

We conclude by discussing some possible short and mid-term improvements of the newly discussed model. 
First, even considering non-evolving hosts with smooth density profiles, the DF description could be improved in multiple ways.
For instance when computing the DF expression for the disc we assumed an isotropic Gaussian as distribution function. A natural extension is to consider an anisotropic version of such distribution function encoding different velocity dispersion along different direction, such as $\sigma_R = \sigma_\phi > \sigma_z$ \citep[see e.g.][]{Binney1977}. Despite this choice inevitably requires the numerical computation of some integrals, the task could likely be optimised in order to maintain good computational efficiency.
A further (indirect) improvement for disc DF consists in the development of more realistic expressions for the rotational pattern $\mathbf{v}_{\rm rot}$ and $\sigma_R$ profile, such that a well-behaved form is maintained also in the central regions. 

A more challenging aspect concerns how the distribution functions of the individual galactic components are evaluated. In principle, such functions should be computed self-consistently considering the whole galaxy potential, while in the current analysis the velocity distributions of the bulge and halo are computed as if they were in isolation. In addition, we follow the approximation discussed in \citet{Chandra43}, in which only stars moving slower than the MP contribute to DF. Such approximation is less and less valid for cored spherical systems \citep{Antonini2012}. While such approximation has a limited impact for Hernquist-like profiles (justifying the agreement we find with numerical simulations), it could  severely affect the orbital decay of MPs in bulges with close to constant density cores \citep{Antonini2012}.\footnote{We stress that the results presented in \citet{Antonini2012} refer to the specific case of spherical stellar systems in quasi-Keplerian potentials, i.e. it is valid only for the final stages of DF within the influence sphere of the host central MBH. While the same distribution function has been used for MPs at much larger separations \citep[e.g.][]{Li2020}, a self-consistent implementation would require the use of the correct distribution function at every scale.} 

We also limited our analysis to MPs with fixed masses, while in general a substantial mass evolution is expected in the case of galaxy mergers. For instance, during the early stages of the inspiral, the decaying MBH is expected to be embedded in a sizable fraction of its original stellar core. The resulting MP is therefore characterised by both an increased effective mass (determining a more efficient DF) and a larger size. However, as the decay proceeds, tidal effects exerted by the host galaxy on the MP tend to gradually erode the residual stellar envelope, until we are left with a ``naked'' MBH \citep[see e.g.][]{Sandor14}. On the other hand, if large reservoirs of gas are available and promptly funnelled toward the secondary nucleus, the decaying MBH could accrete a considerable amount of it and thus gradually increasing its mass as time passes \citep{callegari1, capelo1,CD}.

Additional improvements can be considered relaxing the assumption of smooth/axi-symmetric density distributions. Specifically, when considering the galactic gaseous medium, it has been found that in young galaxies gas can be quite turbulent and even clumpy. This introduces stochastic patterns in the motion of MPs, that depending on the specific gas conditions can significantly alter the otherwise smooth decay \citep{Fiacconi2013,Roskar2015, Tamburello2017, Souza-Lima2017}.
Stocasticity is also typical in marginally Toomre unstable galactic discs, which may lead to the formation of bars and/or spirals. Such structures strongly deviates from axi-symmetry and the torques that they exert on inspiralling MPs can significantly disturb their orbits, in the most extreme situations even scattering them away \citep{Bortolas2020}.

Finally, a more realistic semi-analytical model should also consider the possible time evolution of the host galaxy during the orbital evolution of the MP. If the decay from large separation lasts for a quite long time, of the order of several Gyr, the host galaxy can significantly evolve, in particular at high redshift, where accretion of pristine gas from the cosmic filaments can substantially change the total mass and possibly affect the main galaxy geometry \citep[see e.g. the discussion in][]{Yetli}. Major galaxy mergers can also strongly perturb the galaxy structure and determining radical changes for any MP evolution. Unfortunately, given the very complex physics involved in such mergers, a semi-analytical description is challenging and full numerical simulation should be considered. 

We plan to address the above-mentioned caveats by including them in our semi-analytical framework, to provide a fast and as accurate as possible description of the orbital decay of satellites onto evolving disc galaxies.

%%%%%%%%%%%%%%%%%%%%%%%%%%%%%%%%%%%%%%%%%%%%%%%%%%%%%%%%%%%%%%%%%%%%%%%%%%%%%%%%%%%%%
%%%%%%%%%%%%%%%%%%%%%%%%%%%%%%%%%%%%%%%%%%%%%%%%%%%%%%%%%%%%%%%%%%%%%%%%%%%%%%%%%%%%%
\section*{Acknowledgements}

We thank Jo Bovy for insightful discussion and for his precious help in setting up an exponential disc with ${\rm sech^2}$ vertical profile within \galpy.
Numerical calculations have been made possible through a CINECA-INFN agreement, providing access to resources on GALILEO and MARCONI at CINECA. MD, MB, and AL, acknowledge funding from MIUR under the grant PRIN 2017-MB8AEZ.
AL acknowledges support from the European Research Council No. 740120 `INTERSTELLAR'. 
EB acknowledges support from the \textit{Swiss National Science Foundation} under the Grant 200020\_178949. 

\section*{Data Availability Statement}
The data underlying this article will be shared on reasonable request to the corresponding author.


\bibliographystyle{mnras}
\bibliography{biblio} % if your bibtex file is called biblio.bib

%%%%%%%%%%%%%%%%%%%%%%%%%%%%%%%%%%%%%%%%%%%%%%%%%%
%%%%%%%%%%%%%%%%% APPENDICES %%%%%%%%%%%%%%%%%%%%%

\clearpage
\appendix
%%%%%%%%%%%%%%%%%%%%%%%%%%%%%%%%%%%%%%%%%%%%%%%%%%%%%%%%%%%%%%%%%%%
%%%%%%%%%%%%%%%%%%%%%%%%%%%%%%%%%%%%%%%%%%%%%%%%%%%%%%%%%%%%%%%%%%%
\section{Potential and acceleration for thick exponential discs}
\label{sec:appA}

In this appendix, we detail the derivation leading to the form of potential and accelerations employed in this work. Our derivation follows those presented in \citet{Casertano1983} and \citet{Kuijken1989}, with the exception that we express our results in terms of the Gauss hypergeometric function $_2 F_1(a,b;c;z)$ \citep[see e.g.][for full details; see also Appendix~\ref{sec:appB} for the employed specific implementation]{Abramowitz1972}.

We consider a thick exponential disc with density profile given by:
%%%%%%%%%%%%%%%%
\begin{equation}\label{eq:A_density}
    \rho(R,z) = \rho_{d,0} \ \rho_R(R) \ \rho_z(z),
\end{equation}
where
\begin{align}
    \rho_R(R) &= {\rm e}^{-\alpha R}, \\
    \rho_z(z) &= {\rm sech}^2\left(\dfrac{\beta z}{2}\right) = \cosh^{-2}\left(\dfrac{\beta z}{2}\right),
\end{align}
%%%%%%%%%%%%%%%%
where $\rho_{d,0}$ represents the normalisation constant (see equation~\ref{eq:dens_disk}), $\alpha = 1/R_d$ is the inverse of the scale radius, while $\beta = 2/z_d$ is the inverse of the parameter shaping the vertical profile.

In order to obtain the gravitational potential $\phi$ generated by the above mass distribution, we need to solve the Poisson's equation, i.e.
%%%%%%%%%%%%%%%%
\begin{equation}\label{eq:A_poisson}
    \nabla^2 \phi(R,z) = 4\pi G \ \rho(R,z).
\end{equation}
%%%%%%%%%%%%%%%%
When axial symmetry holds, equation~(\ref{eq:A_poisson}) can be solved in terms of Hankel's transform, defined as
%%%%%%%%%%%%%%%%
\begin{align}\label{eq:HT_direct}
    \tilde{f}(k) &= \int_0^\infty \ud R \ R \ J_{0}(k R) f(R),\\
    f(R) &= \int_0^\infty \ud k \ k \ J_{0}(k R) \tilde{f}(k),
    \label{eq:HT_inverse}
\end{align}
%%%%%%%%%%%%%%%%
where $J_0$ is the 0-th order Bessel function of the first kind, and $f(R)$ is a generic function.

By taking the Hankel's transform of both sides of equation~(\ref{eq:A_poisson}), we obtain the linear non-homogeneous second order differential equation

%%%%%%%%%%%%%%%%
\begin{equation}
    -k^2 \tilde{\phi}(k,z) + \dfrac{\partial^2\tilde{\phi}}{\partial z}(k,z) = 4\pi G \ \tilde{\rho}(k,z),
\end{equation}
%%%%%%%%%%%%%%%%
that, once solved through standard techniques, yields

%%%%%%%%%%%%%%%%
\begin{equation}
    \tilde{\phi}(k,z) = - \dfrac{2\pi G}{k} \int_{-\infty}^{+\infty} \ud \zeta \ {\rm e}^{-k |z-\zeta|} \tilde{\rho}(k,\zeta),
\end{equation}
%%%%%%%%%%%%%%%%
provided that the Hankel transform of the density profile, $\tilde{\rho}(k,z)$, vanishes when $|z| \rightarrow \pm \infty$ (actually our case). Through equation~(\ref{eq:HT_inverse}), the potential  reads

%%%%%%%%%%%%%%%%
\begin{equation}\label{eq:A_phi_gen}
    \phi(R,z) = - 2\pi G \int_0^\infty \ud k J_0(k R) \int_{-\infty}^{+\infty} \ud \zeta \ {\rm e}^{-k |z-\zeta|} \tilde{\rho}(k,\zeta).
\end{equation}
%%%%%%%%%%%%%%%%

Since the density profile in equation~(\ref{eq:A_density}) is factorized in the radial and vertical part, the Hankel transform only affects the former, such that equation~(\ref{eq:A_phi_gen}) reads

%%%%%%%%%%%%%%%%
\begin{align}\label{eq:A_phi_spec}
    \phi(R,z) &= - 2\pi G \rho_{d,0} \int_0^\infty \ud k J_0(k R) \quad \times \nonumber\\
    & \times \int_0^\infty \ud u \ u J_0(uk)\rho_R(u) \int_{-\infty}^{+\infty} \ud \zeta \ {\rm e}^{-k |z-\zeta|} \rho_z(\zeta).
\end{align}
%%%%%%%%%%%%%%%%

The $u$-integral can be evaluated analytically \citep[see e.g.][]{Kuijken1989} and gives

%%%%%%%%%%%%%%%%
\begin{equation}
    \int_0^\infty \ud u \ u J_0(uk) {\rm e}^{-\alpha u} = \dfrac{\alpha}{(\alpha^2+k^2)^{3/2}},
\end{equation}
%%%%%%%%%%%%%%%%
while the $\zeta$-integral needs some additional manipulations. In particular, we can first split the integration domain and change the sign of the integration variable of the $[-\infty,z]$ part, obtaining

%%%%%%%%%%%%%%%%
\begin{align}\label{eq:Ik}
    I_z(k) = \int_{-z}^{+\infty} \ud \zeta \ \dfrac{{\rm e}^{-k (z+\zeta)}}{\cosh^2\left(\frac{\beta \zeta}{2}\right)} + \int_{z}^{+\infty} \ud \zeta \ \dfrac{{\rm e}^{-k (\zeta-z)}}{\cosh^2\left(\frac{\beta \zeta}{2}\right)},
\end{align}
%%%%%%%%%%%%%%%%
that after the variable change $\zeta\pm z = (2/\beta) y$ we reduce to

%%%%%%%%%%%%%%%%
\begin{align}
    I_z(k) = \dfrac{2}{\beta}\left[ \int_{0}^{+\infty} \ud y \ \dfrac{{\rm e}^{-2ky/\beta}}{\cosh^2\left(y-\frac{\beta z}{2}\right)} + \int_{0}^{+\infty} \ud y \ \dfrac{{\rm e}^{-2ky/\beta }}{\cosh^2\left(y+\frac{\beta z}{2}\right)} \right].
\end{align}
%%%%%%%%%%%%%%%%
Finally, we use the symbolic software {\it Mathematica} to express it in a closed form in terms of the Gauss hypergeometric function $_2F_1$, i.e.

%%%%%%%%%%%%%%%%
\begin{align}\label{eq:I_z_k}
    I_z(k) = \dfrac{4}{\beta}\Biggl\{ 1 - \dfrac{k}{k+\beta} &\biggl[ {\rm e}^{-z \beta}
    \prescript{}{2}{F}_1\left(1,1+\frac{k}{\beta};2+\frac{k}{\beta};-{\rm e}^{-z \beta}\right) \nonumber\\
    &+ {\rm e}^{z \beta} \prescript{}{2}{F}_1\left(1,1+\frac{k}{\beta};2+\frac{k}{\beta};-{\rm e}^{z \beta}\right) \biggr] \Biggr\}.
\end{align}
%%%%%%%%%%%%%%%%

Going back to equation~(\ref{eq:A_phi_spec}), the final form of the gravitational potential can be written as

%%%%%%%%%%%%%%%%
\begin{equation}\label{eq:A_phi_fin}
    \phi(R,z) = - 2\pi G \alpha \rho_{d,0} \int_0^\infty \ud k J_0(k R) \dfrac{I_z(k)}{(\alpha^2+k^2)^{3/2}}.
\end{equation}
%%%%%%%%%%%%%%%%

To obtain the radial and vertical accelerations necessary to integrate the equations of motion, we take the (negative) gradient of equation~(\ref{eq:A_phi_fin}), obtaining

%%%%%%%%%%%%%%%%
\begin{align}\label{eq:A_acc_R}
    a_R = -\dfrac{\partial \phi}{\partial R} &= - 2\pi G \alpha \rho_{d,0} \int_0^\infty \ud k k J_1(k R) \dfrac{I_z(k)}{(\alpha^2+k^2)^{3/2}}, \\
    a_z = -\dfrac{\partial \phi}{\partial z} &= - 2\pi G \alpha \rho_{d,0} \int_0^\infty \ud k J_0(k R) \dfrac{-\partial_z I_z(k)}{(\alpha^2+k^2)^{3/2}},
    \label{eq:A_acc_z}
\end{align}
%%%%%%%%%%%%%%%%
where $\partial_z I_z(k)$ is given by

%%%%%%%%%%%%%%%%
\begin{align}\label{eq:I_z_k_der}
    \partial_z I_z(k) = \dfrac{4 k^2 \text{sgn}(z)}{\beta(k+\beta)} &\Biggl[ -{\rm e}^{-z \beta}
    \prescript{}{2}{F}_1\left(1,1+\frac{k}{\beta};2+\frac{k}{\beta};-{\rm e}^{-z \beta}\right) \nonumber\\
    &+ {\rm e}^{z \beta} \prescript{}{2}{F}_1\left(1,1+\frac{k}{\beta};2+\frac{k}{\beta};-{\rm e}^{z \beta}\right) \nonumber\\
    & - \dfrac{k+\beta}{k} \tanh\left(\frac{z \beta}{2}\right)\Biggr].
\end{align}
%%%%%%%%%%%%%%%%

Equations~(\ref{eq:A_phi_fin}--\ref{eq:A_acc_z}) express the gravitational potential, the radial and vertical accelerations in form of 1-dimensional integrals. Unfortunately, no further simplifications are possible and no closed form expressions can be found, therefore those integrals have to be evaluated numerically. Despite the apparent simplicity of dealing with 1-dimensional integrals, the presence of the Bessel functions $J_0$ and $J_1$ can make the task quite challenging given their highly oscillatory behaviour, requiring therefore specific integration schemes to obtain sensible outcomes. Nevertheless, once this task is properly accomplished, it allows us to integrate the orbits with arbitrary initial conditions. Specifically, to numerically integrate such kind of oscillatory functions we employ a special variant of the Double Exponential (DE) rule.\footnote{We refer the interested reader to \citet{Ogata2005} and \citet{Krzysztof2016} for a complete description of the method.}  

In the special case of equatorial motion (i.e. $z=0$), equation~(\ref{eq:I_z_k}) can be simplified, reducing equation~(\ref{eq:Ik}) to

%%%%%%%%%%%%%%%%
\begin{align}
    I_0(k) = 2\int_{0}^{+\infty} \ud \zeta \ \dfrac{{\rm e}^{-k \zeta}}{\cosh^2\left(\frac{\beta \zeta}{2}\right)},
\end{align}
%%%%%%%%%%%%%%%%
that with some manipulations evaluates to \citep[see e.g.][p. 383]{Gradshteyn2007}

%%%%%%%%%%%%%%%%
\begin{align}
    I_0(k) = \dfrac{4}{\beta} \Biggl\{ \dfrac{k}{\beta}\biggl[ \psi\left(\dfrac{k}{2\beta}+\dfrac{1}{2}\right) - \psi\left(\dfrac{k}{2\beta}\right) \biggr] - 1 \Biggr\},
\end{align}
%%%%%%%%%%%%%%%%
where the function $\psi$ is the digamma function \citep[see e.g.][]{Abramowitz1972}.

Finally, since the functions $I_z(k)$, $I_0(k)$, and $\partial_z I_z(k)$ contain addition and subtractions of special functions, roundoff error can represent a serious issue for the numerical evaluation of the integrals when $k\rightarrow +\infty$. To circumvent this possible problem, we can replace the expression of those functions with their asymptotic expansions for $k\rightarrow +\infty$, i.e.

%%%%%%%%%%%%%%%%
\begin{align}\label{eq:A_asymp}
    I_z(k) &= \dfrac{8 \ {\rm e}^{-z \beta}}{(1+{\rm e}^{-z \beta})^2 k} + \mathcal{O}(1/k^3),\\ 
    I_0(k) &= \dfrac{2}{k} - \dfrac{\beta^2}{k^3} + \mathcal{O}(1/k^5),\\
    \partial_z I_z(k) &= \dfrac{8 \beta \ {\rm e}^{-z \beta} (1-{\rm e}^{-z \beta})}{(1+{\rm e}^{-z \beta})^3 \ (k+\beta)} + \mathcal{O}(1/k^3).
\end{align}
%%%%%%%%%%%%%%%%

%%%%%%%%%%%%%%%%%%%%%%%%%%%%%%%%%%%%%%%%%%%%%%%%%%%%%%%%%%%%%%%%%%%
%%%%%%%%%%%%%%%%%%%%%%%%%%%%%%%%%%%%%%%%%%%%%%%%%%%%%%%%%%%%%%%%%%%
\section{Computation of the Gauss hypergeometric function}
\label{sec:appB}

The Gauss hypergeometric function $_2F_1(a,b;c;z)$ is a special function defined for $|z| < 1$ by the power series

%%%%%%%%%%%%%%%%
\begin{equation}\label{eq:hyp_def}
    _2F_1(a,b;c;z) = \dfrac{\Gamma(c)}{\Gamma(a)\Gamma(b)} \sum_{n=0}^\infty \dfrac{\Gamma(a+n)\Gamma(b+n)}{\Gamma(c+n)} \dfrac{z^n}{n!},
\end{equation}
%%%%%%%%%%%%%%%%
and by analytic continuation elsewhere on the complex plane. In the above definition $\Gamma(\cdot)$ represents the gamma function. The Gauss hypergeometric function is very general and contains as special cases several other mathematical functions. A detailed discussion about the properties and relations involving the hypergeometric function is definitely beyond the scope of this work, for which we are only interested in the specific case when the hypergeometric function assumes the form

%%%%%%%%%%%%%%%%
\begin{equation}\label{eq:hyp_spec}
    _2F_1(1,1+y;2+y;z),
\end{equation}
%%%%%%%%%%%%%%%%
with $z \in \mathbb{R}$ always negative and $y \in \mathbb{R}$ and positive.

Several software implementations of the $_2F_1$ function exist, but, at least to our knowledge, none of them is able to provide a sensible evaluation for all possible combinations of $(y,z)$, for which reliable results are usually provided only when $|z|<1$.\footnote{A notable exception is the {\it mpmath} Pyhton package, which unfortunately cannot be efficiently used in our C++ implementation.}
We therefore implemented our own algorithm to evaluate $_2F_1$ for $|z| > 1$, relying instead on the GNU GSL mathematical library for $|z| \leq 1$.

In order to efficiently compute $_2F_1$, we first note that two limiting cases exist, i.e. when $y\rightarrow 0$ and $y\rightarrow +\infty$. In these specific situations, $_2F_1$ assumes the following special values 

%%%%%%%%%%%%%%%%
\begin{align}\label{eq:limit_0}
    _2F_1(1,1;2;z) &= -\dfrac{\ln(1-z)}{z},\\
    _2F_1(1,b;b;z) &= (1-z)^{-1},
    \label{eq:limit_inf}
\end{align}
%%%%%%%%%%%%%%%%
allowing us to express $_2F_1$ as one of the limiting value plus small corrections.

Specialising equation~(\ref{eq:hyp_def}) with parameters in equation~(\ref{eq:hyp_spec}), we get

%%%%%%%%%%%%%%%%
\begin{equation}\label{eq:hyp_spec_ser}
    _2F_1(1,1+y;2+y;z) = (1+y) \sum_{n=0}^\infty \dfrac{z^n}{1+y+n}.
\end{equation}
%%%%%%%%%%%%%%%%
In the case of $y \rightarrow 0$, we can expand the function as

%%%%%%%%%%%%%%%%
\begin{equation}
    _2F_1(1,1+y;2+y;z) = (1+y) \sum_{k=0}^\infty (-1)^k y^k \sum_{n=0}^\infty \dfrac{z^n}{(1+n)^{k+1}},
\end{equation}
%%%%%%%%%%%%%%%%
that, changing the summation index to $m = n+1$, yields

%%%%%%%%%%%%%%%%
\begin{equation}\label{eq:hyp1}
    _2F_1(1,1+y;2+y;z) = \dfrac{1+y}{z} \sum_{k=0}^\infty (-1)^k y^k \sum_{m=1}^\infty \dfrac{z^m}{m^{k+1}}.
\end{equation}
%%%%%%%%%%%%%%%%
where the series over $m$ defines the polylogarithm function \citep{Abramowitz1972} of order $(k+1)$, denoted  $Li_{k+1}(z)$.
Starting from the 1st order of the function $Li_1(z) = -\ln(1-z)/z$, all lower (i.e. the 0-th and negative) orders can be expressed in closed form exploiting the recurrence relation

%%%%%%%%%%%%%%%%
\begin{equation}\label{eq:LI_recur}
    z \dfrac{\partial Li_s}{\partial z}(z) = Li_{s-1}(z).
\end{equation}
%%%%%%%%%%%%%%%%
On the other hand, above the 2nd positive order (included) a closed form cannot be found. Exploiting the polylogarithm function, equation~(\ref{eq:hyp1}) can be written as

%%%%%%%%%%%%%%%%
\begin{equation}
    _2F_1(1,1+y;2+y;z) = -\dfrac{1+y}{z}\ln(1-z) + \dfrac{1+y}{z} \sum_{k=1}^\infty (-1)^k y^k Li_{k+1}(z).
\end{equation}
%%%%%%%%%%%%%%%%
Depending on the value of $y$ we can then truncate the series to match a desired accuracy. Finally, by setting $y=0$ in the above equation, we can check that it correctly simplifies into equation~(\ref{eq:limit_0}).

On the opposite limit, i.e. when $y \rightarrow +\infty$, we follow the very same procedure outlined above, this time expanding the term containing $y$ as

%%%%%%%%%%%%%%%%
\begin{equation}
    \dfrac{1+y}{1+y+n} = 1 + \sum_{l=1}^\infty (-1)^l \dfrac{n (1+n)^{l-1}}{y^l},
\end{equation}
%%%%%%%%%%%%%%%%
that, inserted into equation~(\ref{eq:hyp_spec_ser}), gives

%%%%%%%%%%%%%%%%
\begin{equation}
    _2F_1(1,1+y;2+y;z) = \sum_{n=0}^\infty z^n + \sum_{n=0}^\infty z^n \sum_{l=1}^\infty (-1)^l \ \dfrac{n(1+n)^{l-1}}{y^l}.
\end{equation}
%%%%%%%%%%%%%%%%
The first term involving only powers of $z$ is the series definition for $1/(1-z)$. The second term, instead, can be further expanded exploiting the binomial theorem, leading to

%%%%%%%%%%%%%%%%
\begin{align}
    &_2F_1(1,1+y;2+y;z) =\nonumber\\
    &= (1-z)^{-1} + \sum_{l=1}^\infty \dfrac{(-1)^l}{y^l} \sum_{k=0}^{l-1}\binom{l-1}{k} \sum_{n=1}^\infty n^{k+1} z^n,\nonumber\\
    &= (1-z)^{-1} + \sum_{l=1}^\infty \dfrac{(-1)^l}{y^l} \sum_{k=0}^{l-1}\binom{l-1}{k} \ Li_{-(k+1)}(z),
\end{align}
%%%%%%%%%%%%%%%%
where, in the third line, we recognised the definition of the polylogarithm function.
Since the polylogarithm function appears with negative orders, each of them can be easily evaluated through the recurrence relation in equation~(\ref{eq:LI_recur}).

Finally, in all cases for which the limiting forms for $y\rightarrow 0$ or $y \rightarrow +\infty$ are not suitable, we evaluate the hypergeometric function using its series definition in equation~(\ref{eq:hyp_def}). Formally, this provides a sensible result only when $|z|<1$, but exploiting the linear transformation property of $_2F_1$ \citep{Abramowitz1972} and performing the transformation $z \rightarrow 1/z$, an alternative expression of $_2F_1$ as a function of $1/z$ can be obtained. Once reduced and specialised to the case of interest, it reads

%%%%%%%%%%%%%%%%
\begin{equation}
    _2F_1(1,1+y;2+y;z) = \dfrac{1+y}{z}\sum_{n=0}^\infty \dfrac{1}{(n-y)z^n} - \dfrac{\pi (1+y)}{\sin(\pi y) \ z^{1+y}}.
\end{equation}
%%%%%%%%%%%%%%%%

%%%%%%%%%%%%%%%%%%%%%%%%%%%%%%%%%%%%%%%%%%%%%%%%%%%%%%%%%%%%%%%%%%%
%%%%%%%%%%%%%%%%%%%%%%%%%%%%%%%%%%%%%%%%%%%%%%%%%%%%%%%%%%%%%%%%%%%
\section{Useful expressions for DF with fast moving stars}
\label{sec:appC}

Here, we provide the analytical form of the integral over the variable $V$ that appears in equation~\eqref{eq:DF_general_J}, i.e.

%%%%%%%%%%%%%%%%%%
\begin{equation}\label{eq:C_J}
    J(v_p,v_\star,p_{\rm max}) = \int_{|v_p-v_\star|}^{v_p+v_\star} \ud V \left(1+\dfrac{v_p^2-v_\star^2}{V^2}\right) \ln\left(1+\dfrac{p_{\rm max}^2 V^4}{G^2 m_p^2}\right).
\end{equation}
%%%%%%%%%%%%%%%%%%
where $v_p$ and $v_\star$ are the MP and star velocities, whereas $p_{\rm max}$ is the maximum impact parameter. To compute the integral, we need to consider the two different cases in which $v_p > v_\star$ and $v_p < v_\star$.
For convenience, we express the result in terms of the Gauss hypergeometric function $F(\bullet) = \ _2F_1(1,5/4,9/4,\bullet)$.\footnote{Alternatively, the interested reader can express $_2F_1$ in terms of elementary functions employing known mathematical relations. Moreover, following the steps outlined in \citet{Chandrasekhar1941} 
a more complicated version of the $J$ integral, without neglecting some non-dominant terms, can be derived
to obtain a more general, though cumbersome, expression.}

When $v_p > v_\star$, the integration yields
%%%%%%%%%%%%%%%%%%
\begin{align}\label{eq:C_slow}
    &\dfrac{1}{8 v_\star} \int_{v_p-v_\star}^{v_p+v_\star} \ud V \left(1+\dfrac{v_p^2-v_\star^2}{V^2}\right) \ln\left(1+\dfrac{p_{\rm max}^2 V^4}{G^2 m_p^2}\right) \nonumber\\
    & = \dfrac{1}{20 v_\star}\Biggl[ \nonumber\\
    &   -2\left(\dfrac{G m_p}{p_{\rm max}}\right)^2 \dfrac{v_p+v_\star}{(v_p-v_\star)^4} F(\alpha)
        +2\left(\dfrac{p_{\rm max}}{G m_p}\right)^2 (v_p-v_\star)^5 F(\beta)\nonumber\\
    &   +2\left(\dfrac{G m_p}{p_{\rm max}}\right)^2 \dfrac{v_p-v_\star}{(v_p+v_\star)^4} F(\gamma)
        -2\left(\dfrac{p_{\rm max}}{G m_p}\right)^2 (v_p+v_\star)^5 F(\delta) \nonumber\\
    &   +5v_\star\Biggl( 4 + \ln\Bigl( 1 + \dfrac{p_{\rm max}^4}{G^4 m_p^4}(v_p^2-v_\star^2)^4 + \dfrac{2 p_{\rm max}^2}{G^2 m_p^2} (v^4+6v^2v_\star^2+v_\star^4) \Bigr) \Biggr) \Biggr],
\end{align}
%%%%%%%%%%%%%%%%%%
while, when $v_p < v_\star$, we obtain

%%%%%%%%%%%%%%%%%%
\begin{align}\label{eq:C_fast}
    &\dfrac{1}{8 v_\star} \int_{-v_p+v_\star}^{v_p+v_\star} \ud V \left(1+\dfrac{v_p^2-v_\star^2}{V^2}\right) \ln\left(1+\dfrac{p_{\rm max}^2 V^4}{G^2 m_p^2}\right) \nonumber\\
    & = \dfrac{1}{10 v_\star}\Biggl[ -10 v_p \nonumber\\
    &   +\left(\dfrac{G m_p}{p_{\rm max}}\right)^2 \dfrac{v_p+v_\star}{(v_p-v_\star)^4} F(\alpha)
        -\left(\dfrac{p_{\rm max}}{G m_p}\right)^2 (v_p-v_\star)^5 F(\beta)\nonumber\\
    &   +\left(\dfrac{G m_p}{p_{\rm max}}\right)^2 \dfrac{v_p-v_\star}{(v_p+v_\star)^4} F(\gamma)
        -\left(\dfrac{p_{\rm max}}{G m_p}\right)^2 (v_p+v_\star)^5 F(\delta) \nonumber\\
    &   +\dfrac{5}{2}v_\star \ln\left( \dfrac{1 + \frac{p_{\rm max}^2}{G^2 m_p^2}(v_p+v_\star)^4}{1 + \frac{p_{\rm max}^2}{G^2 m_p^2}(v_p-v_\star)^4} \right) \Biggr],
\end{align}
%%%%%%%%%%%%%%%%%%
where in both expression we have defined
%%%%%%%%%%%%%%%%%%
\begin{align}
    \alpha = -\dfrac{G^2 m_p^2}{p_{\rm max}^2 (v_p-v_\star)^4},\\
    \beta = -\dfrac{p_{\rm max}^2 (v_p-v_\star)^4}{G^2 m_p^2},\\
    \gamma = -\dfrac{G^2 m_p^2}{p_{\rm max}^2 (v_p+v_\star)^4},\\
    \delta = -\dfrac{p_{\rm max}^2 (v_p+v_\star)^4}{G^2 m_p^2}.
\end{align}
%%%%%%%%%%%%%%%%%%

We note that, if we assume the logarithmic term in equation~\eqref{eq:C_fast} to be independent of $V$, then the resulting integral vanishes, and the equation simplifies to the well known result in which only particles moving slower than the MP contribute to DF.

%%%%%%%%%%%%%%%%%%%%%%%%%%%%%%%%%%%%%%%%%%%%%%%%%%%%%%%%%%%%%%%%%%%
%%%%%%%%%%%%%%%%%%%%%%%%%%%%%%%%%%%%%%%%%%%%%%%%%%%%%%%%%%%%%%%%%%%
\section{Additional test cases}
\label{sec:appD}

\begin{figure*}
    \centering
    \includegraphics[width=0.48\textwidth]{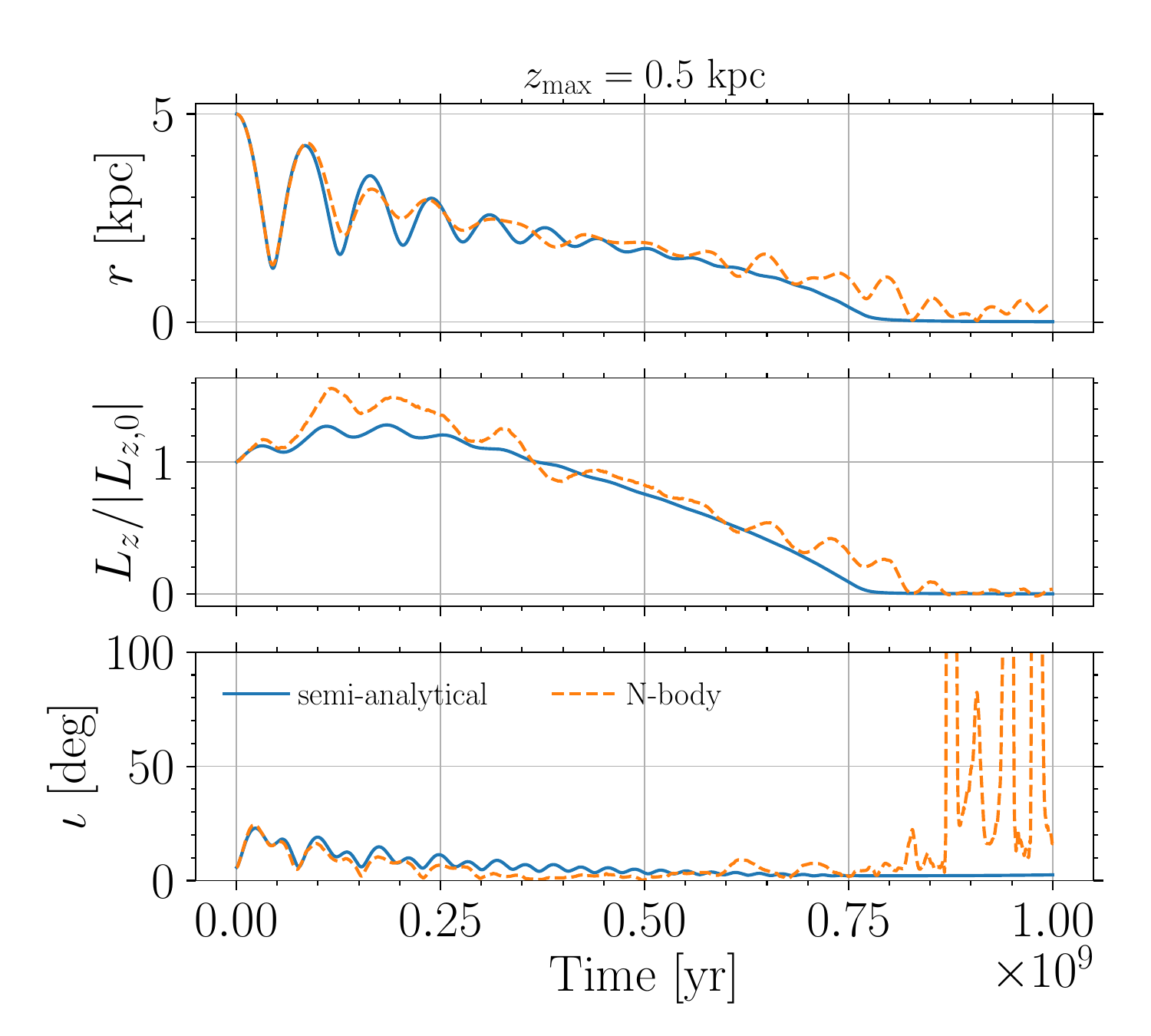}
    \includegraphics[width=0.48\textwidth]{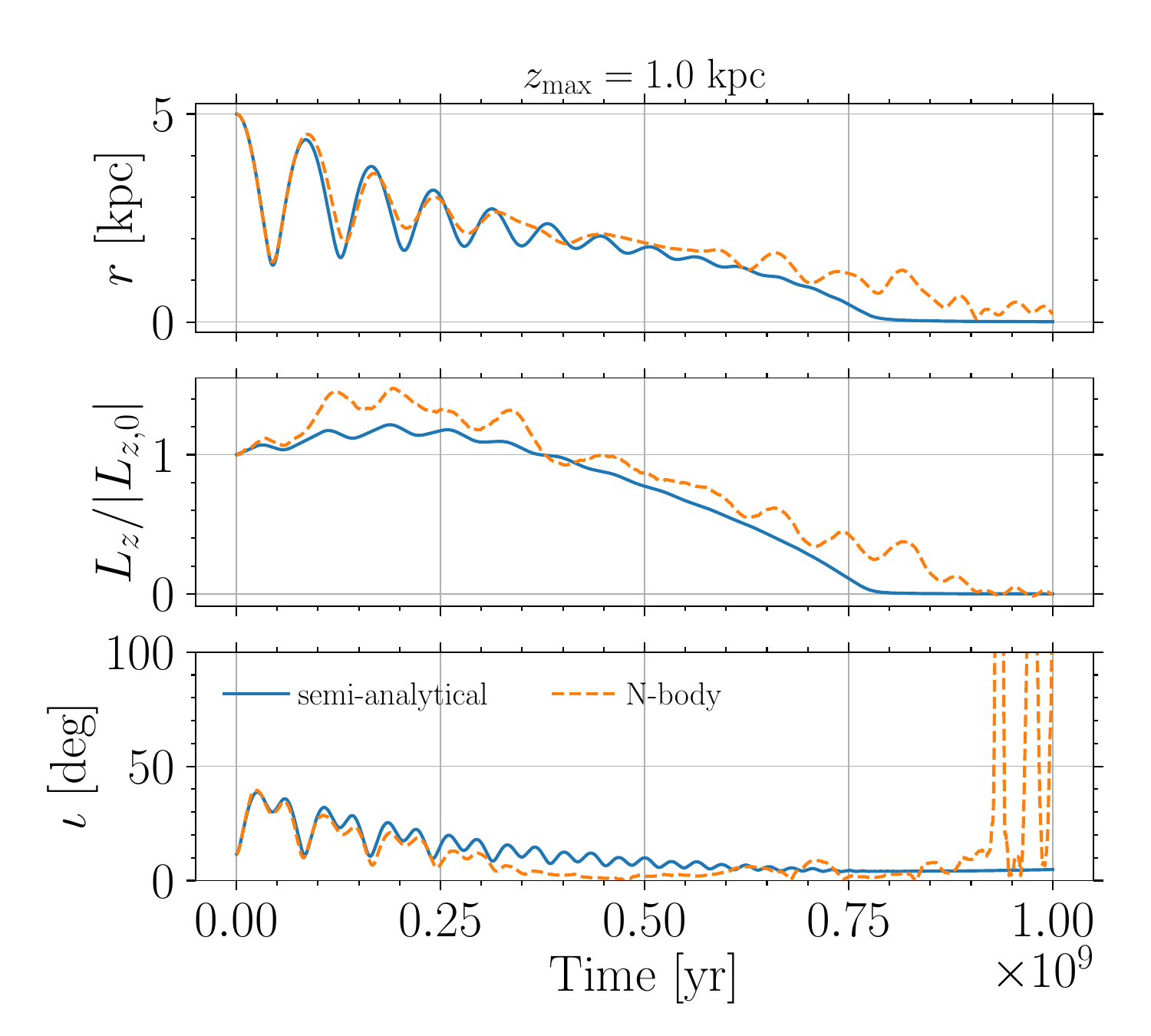}
    \includegraphics[width=0.48\textwidth]{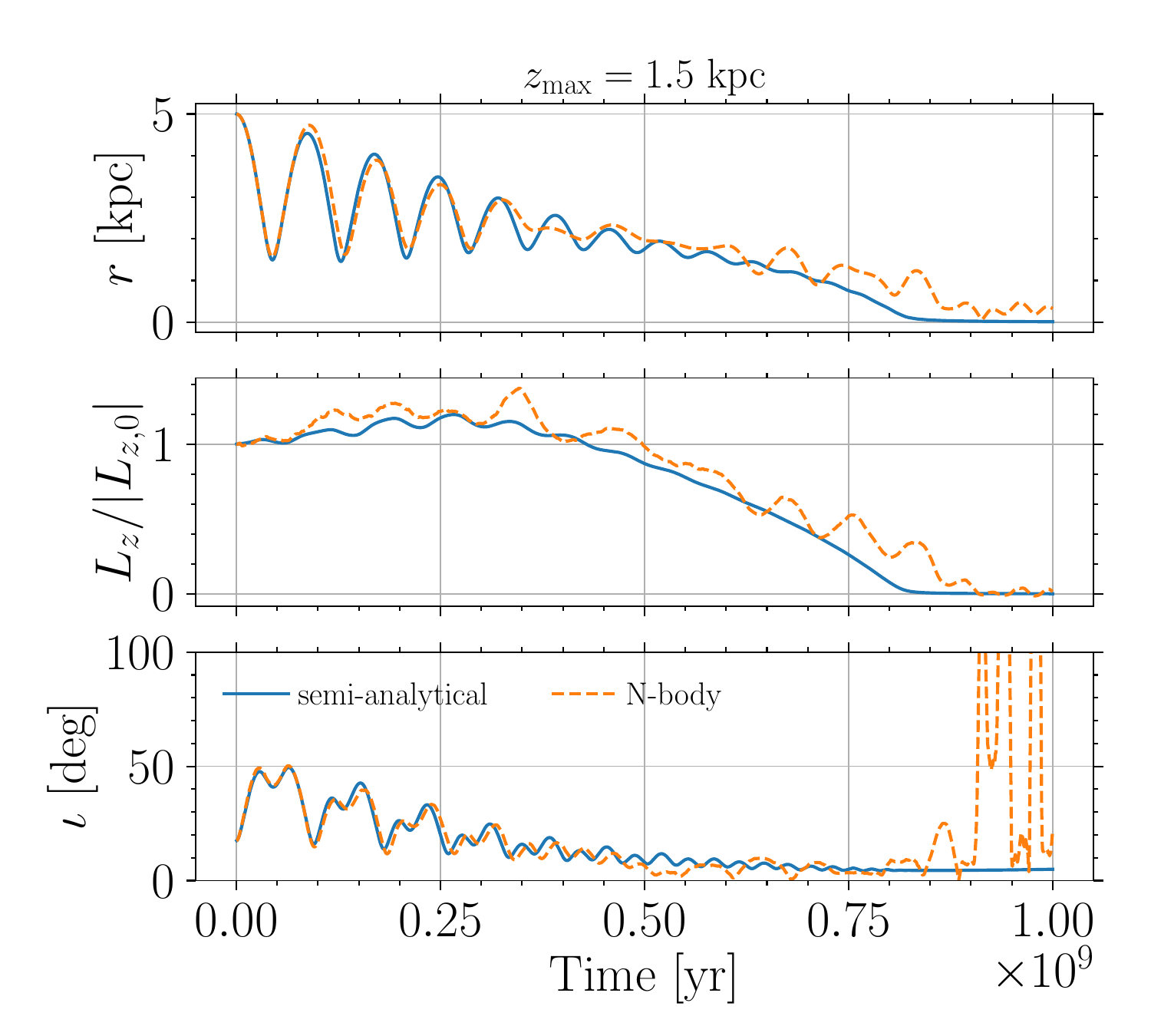}
    \includegraphics[width=0.48\textwidth]{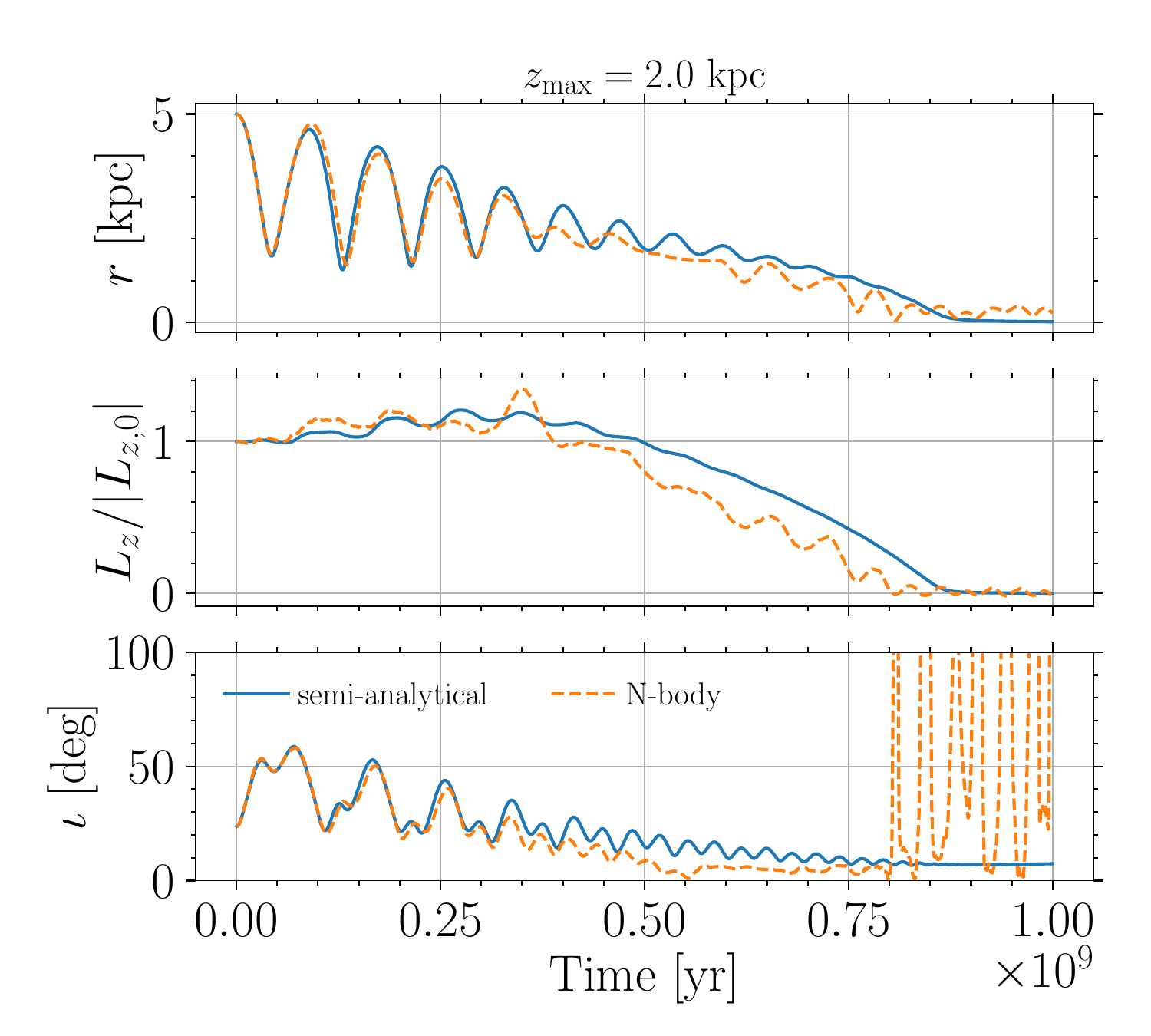}
    \caption{Time evolution of radial separation, $z$ component of the angular momentum (normalised to its staring value) and orbital inclination for four cases featuring an increasing initial $z_{\rm max}$ ranging from 0.5 kpc (top left) to 2 kpc (bottom right). In all plots, the semi-analytical evolution is shown with solid blue lines, while the comparison with the corresponding N-body simulation is given by the dashed orange lines.}
    \label{fig:addtional_cases}
\end{figure*}

Here we show some additional test cases of the evolution computed within our framework and compared to a corresponding N-body simulation. In particular, in Fig.~\ref{fig:addtional_cases} we consider low inclination orbits with the initial $z$ coordinates ranging from 0.5 kpc (top left) to 2.0 kpc (bottom right). For each case we show the time evolution of the radial separation, that of the angular momentum normalised to its staring value and that of the inclination.
As for the cases commented in Section~\ref{sec:results}, we obtain a fairly good agreement when the limited spatial resolution of the N-body simulation is taken into account and also when the tangential velocity profiles is used. As previously noted the semi-analytical evolution reproduces quite well the eccentric inspiral of the MP, with a slightly larger dephasing only when the orbit tends to become more circular. This could hint to the fact that, despite the reasonably good recovered evolution, the assumption of locality made in our framework could be less reasonable as the orbit circularises and that second order effects, connected with the complexity and non-linearity of DF, could start to play a role.

Finally, note that the large variations in inclination that can be observed at later times are  due to spurious noise determined by the sparser particle number in the central regions.

%%%%%%%%%%%%%%%%%%%%%%%%%%%
%%%%%%%%%%%%%%%%%%%%%%%%%%%

%%%%%%%%%%%%%%%%%%%%%%%%%%%%%%%%%%%%%%%%%%%%%%%%%%

% Don't change these lines
\bsp	% typesetting comment
\label{lastpage}
\end{document}